\documentclass[twocolumn]{aastex631}

\newcommand\stardis{\textsc{stardis}}
\usepackage{amsmath}
\usepackage[version=4]{mhchem}

\usepackage{savesym}
\savesymbol{tablenum}
\usepackage{siunitx}
\restoresymbol{SIX}{tablenum}
\usepackage{hyperref}

\shorttitle{\stardis\ Stellar Spectral Synthesis}
\shortauthors{Shields et al.}
\graphicspath{{./}{figures/}}

\usepackage{graphicx}
\usepackage{subfigure}
\usepackage{amsmath}

\usepackage{savesym}
\savesymbol{tablenum}
\usepackage{siunitx}
\restoresymbol{SIX}{tablenum}

\usepackage{CJKutf8}

\begin{document}

\title{Introducing STARDIS: An Open and Modular Stellar Spectral Synthesis Code}

\author[0000-0002-1560-5286]{Joshua V. Shields}
\affiliation{Department of Physics and Astronomy, Michigan State University, East Lansing, MI 48824, USA}
\email{shield90@msu.edu}

\author[0000-0002-0479-7235]{Wolfgang Kerzendorf}
\affiliation{Department of Physics and Astronomy, Michigan State University, East Lansing, MI 48824, USA}
\affiliation{Department of Computational Mathematics, Science, and Engineering, Michigan State University, East Lansing, MI 48824, USA}

\author[0000-0003-0440-3918]{Isaac G. Smith}
\affiliation{Department of Physics and Astronomy, Michigan State University, East Lansing, MI 48824, USA}

\author[0000-0003-4747-4329]{Tiago M. D. Pereira}
\affiliation{Rosseland Centre for Solar Physics, University of Oslo, Oslo, Postboks 1029, 0315, Norway}
\affiliation{Institute of Theoretical Astrophysics, University of Oslo, PO Box 1029, Blindern 0315, Oslo, Norway}

\author[0000-0002-7941-5692]{Christian Vogl}
\affiliation{Max-Planck-Institut f\"ur Astrophysik, Karl-Schwarzschild-Str. 1, 85741 Garching, Germany}

\author[0009-0000-2548-7896]{Ryan Groneck}
\affiliation{Department of Physics and Astronomy, Michigan State University, East Lansing, MI 48824, USA}

\author[0000-0001-7343-1678]{Andrew Fullard}
\affiliation{Department of Computational Mathematics, Science, and Engineering, Michigan State University, East Lansing, MI 48824, USA}

\author[0000-0002-8310-0829]{Jaladh Singhal}
\affiliation{IPAC, California Institute of Technology, Pasadena, CA 91125, USA}

\author[0000-0002-3900-1452]{Jing Lu \begin{CJK*}{UTF8}{gbsn}(陆晶)\end{CJK*}}
\affil{Department of Physics and Astronomy, Michigan State University, East Lansing, MI 48824, USA}

\author[0000-0003-1087-2964]{Christopher J. Fontes}
\affil{Center for Theoretical Astrophysics, Los Alamos National Laboratory, Los Alamos, NM 87545, USA}
\affil{Computational Physics Division, Los Alamos National Laboratory, Los Alamos, NM 87545, USA}

\begin{abstract}
    
We introduce a new 1D stellar spectral synthesis Python code called \stardis. \stardis\ is a modular, open-source radiative transfer code that is capable of spectral synthesis from near-UV to IR for FGK stars. We describe the structure, inputs, features, underlying physics, and assumptions of \stardis\ as well as the radiative transfer scheme implemented. To validate our code, we show spectral comparisons between \stardis\ and \textsc{korg} with the same input atmospheric structure models, and also compare qualitatively to \textsc{phoenix} for solar models. We find that \stardis\ generally agrees well with \textsc{korg} for solar models on the few percent level or better, that the codes can diverge in the ultraviolet, with more extreme differences in cooler stars. \stardis\ can be found at \href{https://github.com/tardis-sn/stardis}{https://github.com/tardis-sn/stardis}, and documentation can be found at \href{https://tardis-sn.github.io/stardis/}{https://tardis-sn.github.io/stardis/}.

\end{abstract}

\section{Introduction} \label{sec:intro}

The practice of measuring chemical abundances in stars has been and remains a powerful tool in astrophysics. This practice can entail comparisons of synthetic stellar spectra to observed ones, specifically matching the observed chemical line features against synthetic spectra generated over a set of astrophysical stellar parameters \citep[i.e., varying the composition and structure of the stellar atmosphere to create a set of models as seen in e.g.,][]{blanco-cuaresma_determining_2014, do_super-solar_2018}. Critical to this process is the generation of synthetic stellar spectra, which has historically been handled by a suite of codes. These include, but are not limited to, \textsc{moog} \citep{sneden_nitrogen_1973, sneden_moog_2012},  \textsc{cmfgen} \citep{hillier_treatment_1998}, \textsc{turbospectrum} \citep{plez_turbospectrum_2012}, \textsc{synthe} \citep{kurucz_synthe_1993, sbordone_atlas_2004}, \textsc{sme} \citep{valenti_spectroscopy_1996}, \textsc{spectrum} \citep{gray_calibration_1994}, \textsc{synspec} \citep{hubeny_synspec_2011}, \textsc{phoenix} \citep{hauschildt_fast_1992, baron_phoenix_2010}, and \textsc{korg} \citep{wheeler_korg_2023}. 
% However, none of the codes are public, open-source, actively maintained, easily accessible and approachable, modular, and written in current popular coding languages. 
% However, no existing code can easily incorporate new physics as needed to simulate the varied stellar atmosphere of unusual stars, such as the surviving companions of supernovae, which can be stripped, chemically non-uniformly polluted, or otherwise perturbed in ways that invalidate the assumptions incorporated into existing stellar codes. 
% Indeed, \textsc{korg} is the only open-source stellar synthesis code released recently or being actively maintained. These problems hinder the continued expansion and application of those codes to meet the evolving needs of the stellar astrophysical community. 

We present \stardis, an open-source finite-difference stellar atmospheric spectral synthesis code written in Python, that is one of the most used and taught languages in  astronomy \citep{greenfield_what_2011, schmidt_numerical_2021}. The foundation of Python as a learning language ensures \stardis\ is approachable and expandable by current and future astrophysicists. To date, no other stellar spectral synthesis Python code exists. \stardis\ is built in part on structures from the \textsc{tardis} code \citep[a supernova spectral synthesis code released in][and the adaptations are detailed explicitly in Section~\ref{sec:code_desc}]{kerzendorf_spectral_2014}. This work uses \stardis\ v2025.04.23,
which functions as a 1D local thermodynamic equilibrium (LTE) spherical and plane-parallel stellar atmospheric radiative transfer code and opacity solver. We have written \stardis\ to be highly modular and easily expandable. Arbitrary chemical compositions, densities, temperatures, etc. can be supplied, allowing for tailored spectral investigations. \stardis\ uses the \textsc{carsus}\footnote{\url{https://tardis-sn.github.io/carsus/}} package to prepare and synthesize novel atomic and molecular data from a variety of sources including National Institute of Standards and Technology (NIST) \citep{NIST_ASD}, the Kurucz database \citep{kurucz_including_2017}, the CHIANTI database \citep{dere_chianti_1997, dere_chiantiatomic_2019}, VALD \citep{piskunov_vald_1995}, and others. 
% both by current active developers, as well as future contributors with highly specific science cases. For instance, non-local thermodynamic equilibrium (NLTE) detailed chemical balance calculations that can be swapped in place of current chemical excitation and level population solvers are in development, and more generally the code is written to be modified and expanded.
Finally, the code is accessible on GitHub and maintained by an active collaboration with detailed documentation for application, maintenance, and extensibility that is intended to provide a low barrier to both usage and contribution.

In Section~\ref{sec:code_desc}, we describe the code including the inputs and the calculations that it performs to obtain a synthetic spectrum. In Section~\ref{sec:code_comp} we compare synthetic spectra produced by \stardis\ primarily to those generated with the same input parameters by \textsc{korg}. We conclude the paper in Section~\ref{sec:stardis_conclusions}. 

\section{Code Description} \label{sec:code_desc}

\begin{figure*}[t] 
    \centering
    \includegraphics[width=.9\textwidth]{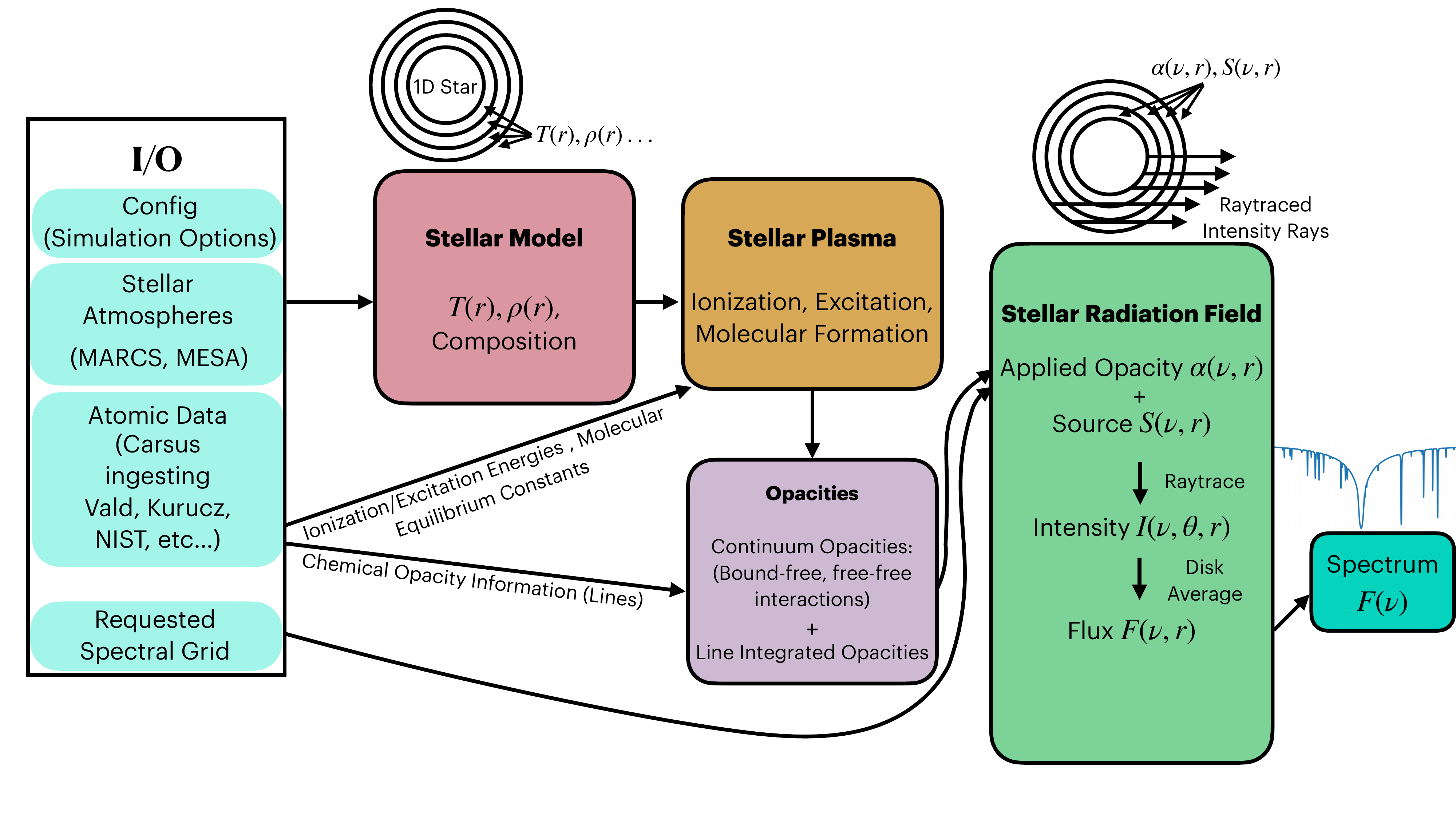}
    \caption{A graphic flowchart showing the operation of \stardis. All inputs are parsed in the input (I/O) stage. The code then creates a \texttt{Stellar Model} that sets up the physical grid at which all future parameters will be evaluated. On this grid, the \texttt{Stellar Plasma} determines the state of the plasma in the atmosphere with the specified quantities, which is then used to calculate the opacity of the plasma. Finally, the simulation raytraces through that plasma a user-specified number of times, and the rays are averaged to create a stellar flux through the atmosphere. The calculated flux that emerges from the top of the atmosphere is  is the output spectrum.}
    \label{fig:code_flowchart}
\end{figure*}

\stardis\ breaks spectral synthesis into four key steps, as shown in Figure~\ref{fig:code_flowchart}. First, it begins in the input stage which ingests the atmosphere, atomic data, and configuration for the code. \stardis\ can currently either use the atmosphere (density, temperature, chemical composition vs depth) output of Model Atmospheres with a Radiative and Convective Scheme (MARCS) \citep{gustafsson_grid_2008} or Modules for Experiments in Astrophysics (MESA) \citep{Paxton2011, Paxton2013, Paxton2015, Paxton2018, Paxton2019, Jermyn2023}, modify one such existing model, or entirely generate a model from scratch based on input from the user. The input module also reads atomic and molecular structure data. The input/output machinery is described in Section \ref{ss:input}.  The next step, molecular formation as well as excitation and ionization of atoms and molecules, is detailed in Section~\ref{ss:chem_eq}. We describe the opacity calculations in Section~\ref{ss:opacities}. The ray-tracing prescription to solve the radiative transfer equation at each depth point throughout the atmosphere is outlined in Section~\ref{ss:raytracing}. Finally, the rays traced through the atmosphere are combined with a weighted average to obtain the disk-averaged stellar spectrum. Note that all calculations, where units are required, are performed in the centimeter-gram-second unit system. 
%This Code Description Section parallels the operation of the code, and will detail each step of the methodology.

\subsection{Inputs} \label{ss:input}

\stardis\ requires four primary inputs for spectral synthesis. First, a chemical data file that includes the necessary information for chemicals in the atmosphere (see Section~\ref{ss:atom_data}). Second, some way to define the stellar atmosphere, notably the structure and composition of (see Section~\ref{ss:atm_ingestion}). Third, a configuration file that specifies the various physics and details of computation (i.e. what physics the user wants to include, as well as computational choices such as how many threads to allow for multiprocessing, how many angles to sample through the atmosphere, etc.). Fourth, the code requires wavelengths or frequencies to compute fluxes on. 

% The atomic data can optionally contain a line list that will be used to calculate line transition features in the spectrum, but not to determine the chemical equilibrium. In addition, it can be supplied with any additional sources of opacity desired by the user, which will usually be sources of continuum opacity not already calculated by the code. Finally, the code requires a configuration file that specifies the exact prescription of code to be run (i.e. what physics the user wants to include, as well as computational choices such as how many threads to allow for multiprocessing, how many angles to sample through the atmosphere, etc.). 

% Note that input models as well as the stellar structure in \stardis\ are evaluated at depth points, not over volume elements. In this way \stardis\ can be referred to as a finite-difference code, rather than finite-volume. Explicitly, it does not consider volume elements of the stellar atmosphere, and instead sampled the stellar atmosphere at specific points and comes up with a description for the atmosphere through differentiation or interpolation between those known points for relevant quantities when necessary.

\subsubsection{Chemical Data and Line Lists} \label{ss:atom_data}

\stardis\ requires atomic weights, ionization energies, energy levels, molecular dissociation energies, and information about line transitions of the atoms and molecules. \stardis\ can read the file format created by the sister package named \textsc{carsus} \citep{passaro_2020_4062427}. As of this work, \stardis\ utilizes \textsc{carsus} synthesized atomic weights, ionization energies, and atomic configurations from NIST, atomic line transitions from \citet{kurucz_atomic_1995}, line transitions from the Vienna Atomic Line Database (VALD) \citep{piskunov_vald_1995, ryabchikova_vienna_1997, kupka_vald-2_1999, kupka_vald-2_2000, ryabchikova_major_2015, pakhomov_evolution_2019}, and molecular dissociation energies, partition functions and equilibrium constants from \citet{barklem_partition_2016}. Once data from these sources are synthesized, however the user specifies, the synthesized data are put into an atomic data file which is required and used by \stardis. 

% Additionally, \textsc{carsus} can parse and prepare a line list, principally from the Vienna Atomic Line Database \citep[VALD][]{piskunov_vald_1995, ryabchikova_vienna_1997, kupka_vald-2_1999, kupka_vald-2_2000, ryabchikova_major_2015, pakhomov_evolution_2019}. When provided, \stardis\ will read and use individual line transitions from these lists. 

\subsection{Stellar Atmosphere Ingestion (\texttt{Stellar Model})} \label{ss:atm_ingestion}

\stardis\ has three ways of defining an atmosphere with specified temperature, density, composition, and geometry on which to solve plasma properties and ultimately perform radiative transfer through. First, \stardis\ can read output files of either MARCS\footnote{Available at \href{https://marcs.oreme.org/}{MARCS}} \citep{gustafsson_grid_2008} or MESA \citep{Paxton2011, Paxton2013, Paxton2015, Paxton2018, Paxton2019, Jermyn2023}\footnote{See the \href{https://docs.mesastar.org/en/release-r23.05.1/index.html}{MESA homepage}}. Second, a user can read a file similar to the previous option, but modify any of the physical properties as desired. Finally, a user can define a stellar atmosphere by hand, which is particularly useful for unique analysis of exotic atmospheres or toy model exploration.

The output of this step is a \texttt{Stellar Model} object which holds the temperatures, densities, geometry, and composition of the atmosphere. The object primarily exists to hold and pass the necessary information downstream to the \texttt{Stellar Plasma} and \texttt{Stellar Radiation Field} objects. Specifically, the geometry determines if the system is treated as  a plane-parallel or spherical geometry as well as the distance between atmospheric grid  points.

% This is particularly a danger to MESA models, where an atmosphere may not be sufficiently defined to perform accurate radiative transfer, either as a result of too low spatial resolution or not solved out to the extended atmosphere. 

\subsection{Plasma Equilibrium Density Calculations (\texttt{Stellar Plasma})} \label{ss:stellar_plasma}

In order to calculate the opacity of the atmosphere, the code must first understand the specific states of the atoms and molecules within the plasma. The \texttt{Stellar Plasma} handles this part of the computation, and exists to solve the molecular, ionization, and excitation balances for each cell in the atmosphere. The resulting number densities are used to calculate atmospheric opacities in Section~\ref{ss:opacities} necessary for the ray-tracing step. 

% The \texttt{Stellar Plasma} object exists to solve and hold the properties of plasma that ultimately lead to the determination of the opacity of the plasma. All of the equations following in this section are solved at each stellar atmosphere point specified by the \texttt{Stellar Model}. 
% The \stardis\ \texttt{Stellar Plasma} object is built as an extension of the \textsc{tardis} plasma, and further detailed description of it can be found ...

\subsubsection{Thermodynamic Equilibrium} \label{ss:chem_eq}

Atomic, ionic, and level populations are calculated under the assumption of LTE using code modules from \textsc{tardis}. First, to calculate the ionization states of the atoms and molecules, we use the Saha ionization equation,

\begin{equation} \label{eq:saha}
    \frac{n_{i, j+1}n_\textrm{e}}{n_{i,j}} = \frac{2Z_{i,j+1}(T)}{Z_{i,j}(T)} \left(\frac{2 \pi m_\textrm{e} k_\textrm{B} T}{h^2} \right)^{3/2} e^{-\chi_{i,j}/k_\textrm{B} T}
\end{equation}
where $n$ is the number density, $i$ and $j$ are indices for element and ionization state, $Z$ is the partition function, $\chi$ is the ionization potential, $m_\textrm{e}$ is the mass of an electron, $k_\textrm{B}$ is the Boltzmann constant, and $T$ is the temperature. Equation~\ref{eq:saha} is solved in \stardis\ as a system of equations where $n_e$ is a shared variable across the set of ionization balance equations.

Second, we use the Boltzmann equation to obtain the populations (or number densities) of the excited states,

\begin{equation} \label{eq:boltzmann}
    n_{i,j,k} = \frac{g_{i,j,k}}{Z_{i,j}}n_{i,j}e^{-\epsilon_{i,j,k} / k_\textrm{B} T}
\end{equation}
 where, in addition to the previous variables, we introduce $k$ as an excited-state index, $g$ as the degeneracy of the state, and $\epsilon$ as the excitation energy of the state relative to the ion ground state.

\subsubsection{Molecular Equilibrium} \label{ss:molecules}

To calculate the presence of molecules in the stellar atmosphere, \stardis\ calculates molecular number densities using equilibrium constants included in the atomic data via \textsc{carsus} for diatomic molecules (see Section~\ref{ss:atom_data}). Equilibrium constants are currently provided from  \citet{barklem_partition_2016}, although constants can also be modified by the user. In this version of \stardis, molecules are not considered for ionization or recombination after formation, and instead form directly into the appropriate ionization state of the molecule (e.g., {H}$_{2}^+$ is formed directly from \ion{H}{1} and \ion{H}{2}). %Maybe this is where I want a note describing the assumption of low molecule limits?
Equilibrium constants are interpolated and resampled to the temperatures of each location of the \texttt{Stellar Plasma}. Molecular number densities are then calculated following: 

\begin{equation}
    n_{\textrm{mol}} = \frac{n_{i,j}^a n_{i,j}^b}{K(T)}
\end{equation}
where $n_{i,j}^a$ and $n_{i,j}^b$ indicate the number density of the constituent ions of the molecule, and $K(T)$ is a number density equilibrium constant. 

\subsection{Opacities} \label{ss:opacities}

\stardis\ calculates each opacity by source independently before summing them together for each wavelength or frequency point given within a simulation.

Each opacity calculation follows the same scheme of calculating a cross section of a particle at a specific frequency due to some physical process and then multiplying that cross section by the number density of that particle. That is,

\begin{equation} \label{eq:opacity}
    \alpha_{i,j}(\nu) = \sigma_{i,j}(\nu) n_i
\end{equation}
where $\sigma$ is the cross section, and $n$ is the number density and $\alpha$ is the attenuation coefficient and is the desired quantity of this Section for each source to be used later in Sections~\ref{ss:applied_opacities} and \ref{ss:raytracing}. $i$ now indexes the particle and $j$ indexes the physical process that causes the particle to contribute to the opacity. Cross sections, and thus opacities, of specific particles and processes are often, but not always, dependent on the frequency $\nu$. Particle number densities are calculated previously as detailed in Section~\ref{ss:stellar_plasma}.

Opacity sources can be largely divided into two categories, i.e., continuum opacity sources and line opacity sources. 

\subsubsection{Continuum Opacities} \label{ss:cont_opa}

Continuum opacities are broadly defined as any opacity source that contributes to a large range of frequencies or wavelengths. \stardis\ considers four continuum opacity sources: Thomson scattering, Rayleigh scattering, bound-free interactions, and free-free interactions. 

\paragraph{Thomson scattering} Thomson scattering is the most simple opacity source to model as it is independent of frequency and only due to a single particle type (i.e., electrons). With $\sigma_{\textrm{T}}$ as the Thomson scattering cross section, Equation \ref{eq:opacity} yields

\begin{equation}
    \alpha_{\textrm{T}} = \sigma_{\textrm{T}}  n_\textrm{e}
\end{equation}
where $n_\textrm{e}$ is the electron density. The following equations in this section will only show cross sections, but the attenuation coefficient calculation is the same. 

\paragraph{Rayleigh scattering} We implement Rayleigh scattering following approximations given in \citet{colgan_new_2016} for the H and He cross sections, relative to the Thomson scattering cross section: 

\begin{equation}
    \sigma_{\textrm{H }{\rm I}} / \sigma_\textrm{T} = 20.24 \left(\frac{\hbar \nu}{2E_\textrm{H}}\right)^4 + 239.2  \left(\frac{\hbar \nu}{2E_\textrm{H}}\right)^6 + 2256 \left(\frac{\hbar \nu}{2E_\textrm{H}}\right)^8 
\end{equation}

\begin{equation}
    \sigma_\textrm{He {\rm I}} / \sigma_\textrm{T} = 1.913 \left(\frac{\hbar \nu}{2E_\textrm{H}}\right)^4 + 4.52  \left(\frac{\hbar \nu}{2E_\textrm{H}}\right)^6 + 7.90 \left(\frac{\hbar \nu}{2E_\textrm{H}}\right)^8  
\end{equation}
where $E_\textrm{H}$ is the Rydberg energy.

\paragraph{Bound-free and free-free opacity} Both bound-free and free-free cross sections are either calculated using an analytic approximation or sampled from an interpolated function constructed from empirically measured opacity tables. Our analytic approximations are implemented following \citet{hubeny_theory_2014}. For bound-free transitions of hydrogenic species, the cross section is

\begin{equation} \label{eq:bf}
    \sigma_{\textrm{bf}} = \frac{64 \pi^4\textbf{Z}^4 e^{10} m_\textrm{e}}{3 \sqrt{3} c h^6} \frac{\bar{g}_{\textrm{bf}}}{n'^5\nu^3}
\end{equation}
where $\textbf{Z}$ is the charge of the atom or molecule, $\bar{g}_\textrm{{bf}}$ is the bound-free Gaunt factor, $n'$ is the principal quantum number, and $\nu$ is the ionization frequency. Gaunt factors represent quantum mechanical departures from the classical cross sections and are measured empirically and appear here as linear scaling to the uncorrected cross sections. 

Similarly, the free-free absorption cross section is given by

\begin{equation}\label{eq:ff}
    \sigma_\textrm{{ff}} = \frac{\sqrt{32\pi} \textbf{Z}^2 e^{6}}{3 \sqrt{3} c h (k_\textrm{B} m_e^3 T)^{1/2}} \frac{\bar{g}_\textrm{{ff}}}{\nu^3}
\end{equation}
where $\bar{g}_\textrm{{ff}}$ is again a Gaunt factor correcting the equation for quantum mechanical effects, specific to the free-free interaction. 

These analytic approximations allow \stardis\ to include the bound-free and free-free opacities for any generic atom or molecule as needed. Alternatively, \stardis\ can use empirically measured cross sections. When requested, \stardis\ linearly interpolates a custom cross section table over wavelengths, or wavelengths and temperatures, as appropriate, and resamples that table at each point in the \texttt{Stellar Plasma}. Once obtained, the opacities are calculated in the usual way following Equation~\ref{eq:opacity}. The currently implemented custom opacity sources are: \ce{H-} bound-free interactions from \citet{wishart_bound-free_1979} (i.e., the interaction of a photon with a proton with two bound electrons, which is a large source of continuum opacity in the solar atmosphere), \ce{H-} free-free interactions\footnote{Note that in bound-free interactions, the particle name of the interaction is given by particle including the free electron, while in free-free interactions the opposite is true. E.g., a \ce{H} nucleus with two bound electrons is responsible for \ce{H-} bound-free interactions, while \ion{H}{1} is responsible for \ce{H-} free-free interactions.} from \citet{bell_free-free_1987}, and \ce{H2+} bound-free interactions following \citet{stancil_continuous_1994}.

\subsubsection{Integrated Line Opacities} \label{ss:lines}

To calculate bound-bound (i.e., line) transition opacities, \stardis~follows \citet{rybicki_radiative_1986}.  Beginning from Equation~\ref{eq:opacity}, we substitute for the frequency integrated cross section of an atom for an excitation transition from a lower state to an upper state which is given by

\begin{equation}
    \sigma_\textrm{line} = \frac{\pi e^2}{m_\textrm{e} c} f_{lu}
\end{equation}
noting that the subscript $l$ denotes the lower level, the subscript $u$ denotes the upper level of a transition, and $f_{lu}$ is the oscillator strength associated with the transition from $l$ to $u$. We calculate the line specific attenuation coefficient $\alpha_\textrm{line}(\nu)$ assuming complete redistribution with

\begin{equation} \label{eq:alpha_line}
    \alpha_\textrm{line}(\nu) = \frac{\pi e^2}{m_\textrm{e} c} n_{l} f_{lu}\Big[1-\dfrac{g_{l}n_{u}}{g_{u}n_{l}}\Big] \phi(\nu).
\end{equation}
$n$ continues to refer to number density, while $g$ refers to the degeneracy of the relevant excitation state. The $[1-\dfrac{g_{l}n_{u}}{g_{u}n_{l}}]$ term appears as the correction for stimulated emission and accounts for a reduction in opacity related to photons being re-emitted at the same frequencies that are being absorbed \citep{rybicki_radiative_1986}. $\phi(\nu)$ is the line profile function, which handles the broadening of the line across frequency space and is discussed in Section~\ref{ss:broadening}.

Before proceeding, we must discuss a necessary implementation detail. Equation \ref{eq:alpha_line} is general, but cannot always be used when incomplete information about a line transition is provided, as can be the case when \stardis\ calculates line opacities using data from line lists.
Line lists consist of line transitions and associated wavelengths, oscillator strengths, or combined oscillator strength and degeneracies (often referred given as $\log gf$ values) for those transitions. 
% In contrast, a line list does not attempt to fully reconstruct an atom and supply an excitation energy and degeneracy for every energy level of an atom, and an oscillator strength of each possible transition. Rather, a line list will be provided for a given wavelength or frequency regime, and specific line transitions will be listed in that region. While not necessarily physically consistent, line lists allow for empirical calibration and tuning to produce more accurate individual transitions than are currently possible from first principles. In practice, this means that a line list may list a specific transition without all of the associated degeneracies, and instead report a single "gf" value. 
Because degeneracies are not always explicitly listed for both states in a line transition, we cannot always solve for the previous stimulated emission factor explicitly. In this case, we modify the stimulated emission factor to $[1-e^{-h \nu / k_B T}]$, using the LTE dependent equivalence

\begin{equation} \label{eq:LTE_stimulated_emission}
\frac{n_l}{n_u} = \frac{g_l}{g_u} e^{h
\nu/k_\textrm{B}T}.
\end{equation}
from \citet{rybicki_radiative_1986}.

% the level populations associated with the transition are not explicitly calculated, and the code instead calculates an opacity of the form 

% \begin{equation} \label{eq:alpha_line_short}
%     \alpha_{\textrm{linelist}}(\nu) = \dfrac{\pi e^{2}}{m_{\textrm{e}} c} n_{l} f_{lu} 
%         \Big[1-e^{-h \nu / k_\textrm{B} T} \Big]\phi(\nu)
% \end{equation}

% where the stimulated emission factor is now $[1-e^{-h \nu / k_B T}]$ and is adopted from \cite{rybicki_radiative_1986}. 

% \stardis\ can ingest line lists generated by the VALD database \citep{piskunov_vald_1995, ryabchikova_vienna_1997, kupka_vald-2_1999, kupka_vald-2_2000, ryabchikova_major_2015, pakhomov_evolution_2019} which includes individual lines from a wide variety of sources. %NOTE come back to this 

\subsubsection{Line Broadening} \label{ss:broadening}

The opacity contribution of an atomic line feature is important over a range of frequencies. This effect is captured by the appearance of the line profile function, $\phi(\nu)$, at the end of Equation~\ref{eq:alpha_line}. The shape of $\phi(\nu)$ is influenced by many physical processes, but in this version of \stardis\ it is taken to be a pure Voigt profile \citep{Voigt1912}, which is the convolution of a Gaussian and a Lorentzian distribution. The Gaussian distribution is the result of Doppler shifts from the thermal and non-thermal motion of particles, while the Lorentzian distribution is the result of every other physical broadening source. The convolution is given by:   
% The Voigt profile for a transition is centered on the frequency that a photon would have with the same energy as the difference between the energies of the two atomic states associated with the line transition. 
% To calculate frequency dependent line opacities, calculating a Voigt profile for each transition is an important step of the \stardis\ code: 

% do not operate at discrete infinitely thin wavelengths because the features are subject to line broadening. This broadening due to thermal motion of the plasma as well as non-thermal physical processes. The broadening of a line due to thermal motion is well described by a Gaussian distribution, while the non-thermal motions are described by a Lorentzian distribution. The two distributions convolved together are known as a Voigt profile, and calculating the Voigt profile is a necessary component of line opacity.  A line transition can be prompted by a photon over a range of frequencies with varying probabilities that can be described by a Voigt profile. A Voigt profile is the convolution of a Gaussian and a Lorentzian, which describe the thermal broadening and the non-thermal sources of line broadening respectively. 

\begin{equation}
    \phi(x; \sigma,\gamma) = \int_{-\infty}^{\infty} G(x';\sigma)L(x-x';\gamma)dx'
\end{equation}
where $G(x;\sigma)$ is the expected Gaussian

\begin{equation}
    G(x;\sigma) = \frac{1}{\sigma \sqrt{2\pi}}\textrm{exp}\left(-\frac{x^2}{2\sigma^2}\right)
\end{equation}
where $\sigma$ is the standard deviation of the Gaussian, and the Lorentzian $L(x;\gamma)$ is written as

\begin{equation}
    L(x;\gamma) = \frac{(\gamma/\pi)}{x^2 + \gamma^2}
\end{equation}
where $\gamma$ describes the broadness of the profile which is also referred to as the half-width at half-maximum of the function. No closed-form solution to the integral Voigt profile exists, however, it can be evaluated instead by

\begin{equation}
    \phi(x; \sigma, \gamma) = \frac{Re[w(z)]}{\sigma\sqrt{2\pi}}
\end{equation}

\begin{equation}
    z = \frac{x + i\gamma}{\sigma\sqrt{2}}
\end{equation}
where $w$ is the Faddeeva function 

\begin{equation}
    w(z) = e^{-z^2} \left(1 + \frac{2i}{\sqrt{\pi}}\int_0^ze^{t^2}dt\right)
\end{equation}
In effect, with an evaluated Gaussian $\sigma$ and Lorentzian $\gamma$, the Voigt profile for any given is specified.

The Gaussian component encompasses two physical effects: 1) thermal motion of atoms and molecules and 2) a parameter often referred to as a microturbulent velocity that is introduced to account for bulk motions of the stellar atmosphere due to convection or other motion on scales smaller than the mean free path of a photon:

\begin{equation}
    \sigma = \frac{\nu}{c} \sqrt{\frac{2k_\textrm{B} T} {m} + \xi^2}
\end{equation}
where $\sigma$ is the spread due to Brownian motion and $\xi$ is a microturbulent velocity parameter. 

The reader should note that $\xi$ is an empirical approximation to give better empirical fits to stellar observations \citep{de_jager_high-energy_1954, mucciarelli_microturbulent_2011}, although both its importance and physical accuracy are debated \citep[see e.g.,][]{asplund_new_2005}. 
% Often codes will modify this equation to include an additional microturbulence parameter $\xi$

% \begin{equation}
%     \sigma = \frac{\nu}{c} \sqrt{\frac{2k_BT} {m} + \xi^2}
% \end{equation}

% We have chosen to neglect this for now. Microturbulence lacks physical motivation and whether or not it improves the accuracy of lines is still in question \citep[see e.g.][]{asplund_new_2005, kuperus_finding_2022}. \stardis\ may include microturbulence in the future specifically to enable better comparisons to other codes, but are somewhat unsure of its scientific justification. 

The width of the Lorentzian in the Voigt profile is given by the summation of three independent components

\begin{equation} \label{eq:gamma_tot}
    \gamma_\textrm{{total}} = \gamma_{\textrm{Stark}} + \gamma_{\textrm{vdW}} + \gamma_{\textrm{rad}},
\end{equation}
where $\gamma_{\textrm{Stark}}$ is responsible for Stark Broadening, $\gamma_{\textrm{vdW}}$ is responsible for van der Waals broadening, and $\gamma_{\textrm{rad}}$ is responsible for radiation or natural broadening. 

\paragraph{Stark effect} The Stark effect describes the splitting of transition lines that are superpositions of transitions between multiple degenerate atomic states into distinct transitions due to the presence of electric fields produced by neighboring electrons and ions in the plasma. Stark broadening is routinely described as an expansion where the linear term describes the leading monopole term and the quadratic term describes the leading dipole term. \ion{H}{1} and hydrogenic particles are thus particles in which linear Stark broadening is important due to their significant polarity, whereas quadratic Stark broadening is important for all other lines. \stardis\ currently implements linear stark broadening solely for H following \cite{sutton_approximate_1978}

\begin{equation}
    \gamma_\textrm{{Stark, linear}} = a_1[0.6 \cdot (\mathbf{n}_u^2 - \mathbf{n}_l^2)n_\textrm{e}^{2/3}]
\end{equation}
%
%Maybe c dot instead of multiplication
% should not be italicized for non variables
where $a_1 = 0.642$ for the first line in a series and 1 otherwise, $\mathbf{n}_u$ and $\mathbf{n}_l$ are the principal quantum numbers of the upper and lower states in the transition, and $n_e$ is again the electron number density. 
% Note that for this Section n will to refer to the quantum number, while $n$ refers to number density as before.

For all non-hydrogen lines, we calculate quadratic Stark broadening following \cite{gray_observation_2005}

\begin{equation}
    \log\gamma_4 = 19 + \frac{2}{3}\log C_4 + \log P_e - \frac{5}{6}\log T
\end{equation}
where $P_\textrm{e}$ is the electron pressure following the ideal gas law to get $P_\textrm{e} = n_\textrm{e} k_\textrm{B} T$. $C_4$ is calculated following \citet{traving_buchbesprechung_1960}

\begin{equation}
    C_4 = \frac{e^2 a_0^3}{36 h \epsilon_0 Z^4} (5 \mathrm{n}_{\textrm{eff},u}^3 + \mathrm{n}_{\textrm{eff},u} - 5 \mathrm{n}_{\textrm{eff},l}^3 - \mathrm{n}_{\textrm{eff},l}).
\end{equation}
In this equation, $e$ is the elementary charge, $Z$ is the ion number, $\epsilon_0$ is the vacuum permittivity of free space, and $\mathrm{n}_{\textrm{eff}}$ is the effective principal quantum number given by

\begin{equation}
    \mathrm{n}_{\textrm{eff}} = \sqrt{\frac{E_\textrm{H}}{\chi - \epsilon}} Z,
\end{equation}
approximating a particle as hydrogenic with a modified electric potential. $E_\textrm{H}$ refers to the Rydberg energy, while $\chi$ refers to the energy required to ionize an electron from the ground state to its current ionization state, and $\chi$ refers to the energy required to excite the electron from the ground ionization state to its current excited ionization state.

\paragraph{Van der Waals effect} Van der Waals broadening, or pressure broadening, is the result of inter-atomic forces on particles involved in line transitions in stellar atmospheres, chiefly a result of \ion{H}{1} applying attractive and repulsive forces at small distances due to the particle's polarization. Van der Waals broadening is the dominant source of broadening for most lines in dwarf atmospheres. This broadening term scales with the number density of \ion{H}{1}. \stardis\ calculates the van der Waals contribution to the Lorentzian of the Voigt profile following \citet{warner_effects_1967}:

\begin{equation}
    \gamma_{\textrm{vdW}} = 17 \left(\frac{8k_\textrm{B}T}{\pi m_\textrm{H}}\right)^{0.3} C_6^{0.4} n_\textrm{H \rm{I}}
\end{equation}
where 
\begin{equation}
    C_6 = 6.46 \cdot 10^{-34} (5 \mathrm{n}_{\textrm{eff},u}^4 + \mathrm{n}_{\textrm{eff},u}^2 - 5 \mathrm{n}_{\textrm{eff},l}^4 - \mathrm{n}_{\textrm{eff},l}^2).
\end{equation}
In these equations, $m_\textrm{H}$ is the mass of H, and $n_\textrm{H \rm{I}}$ is the number density of \ion{H}{1}.

\paragraph{Radiation broadening}\ The radiation broadening (also sometimes referred to as natural broadening) term is a result of the uncertainty in the energy of a particle described by the energy-time uncertainty principle. The radiation broadening term can be approximated directly as the Einstein spontaneous emission coefficient $A_{ul}$. 

Alternatively, each of the broadening parameters discussed in this section are sometimes estimated empirically on a per line basis \citep[see][]{piskunov_vald_1995, anstee_width_1995, barklem_broadening_1998}. When provided by VALD line lists and requested in the configuration, \stardis\ will scale the provided line by line broadening parameters to the properties of the stellar atmosphere where those parameters are evaluated (usually with some power law temperature, or temperature and density dependence), instead of using the equations given earlier in this section.

To obtain the Voigt profile for a given line, \stardis\ follows the numerical approximation of the Voigt profile of \citet{humlicek_optimized_1982}. Through testing, we found that this implementation agrees with the \textsc{SciPy} \citep{virtanen_scipy_2020} numerical implementation of the Voigt profile within $0.01\%$ and is compatible with \textsc{numba} \citep{lam_numba_2015}, discussed further in Appendix~\ref{sec:benchmarks}.

\subsection{Radiation Transport (\texttt{Stellar Radiation Field})} \label{ss:stellar_radiation_field}

The \texttt{Stellar Radiation Field} object exists to solve and then hold the characteristics of the radiation field at different angles and depth points in the stellar atmosphere, at specific requested spectral points. Up to this point in the simulation, all calculations have been performed agnostic to specific frequencies because the opacities have not yet been evaluated at any given wavelength or frequency. At this stage, a user-defined spectral grid enters the simulation, the opacities are applied to the spectral grid, and rays are traced through the atmosphere on that spectral grid. 

\subsubsection{Applied Opacities} \label{ss:applied_opacities}

Before the radiative transfer equation can be solved, and the radiation field through the atmosphere computed at each frequency, the code requires a measurement of the total opacity at each frequency and point in the atmosphere. This is done by applying each opacity source to each requested spectral grid point, evaluated at the appropriate properties of the plasma at that point,

\begin{equation} \label{eq:sum_opacity}
    \alpha_\textrm{total}(\nu, r) = \sum_{i,j} \alpha_{i,j}(\nu, r)
\end{equation}
where $\alpha$ is the opacity from Equation~\ref{eq:opacity}, $\nu$ is a frequency, $r$ is the stellar radius, $i$ indexes over the opacity source particle, $j$ indexes over the ways that the particle contributes to opacity (e.g., bound-free or line). For continuum opacity sources, this entails adding the evaluated equations from Section~\ref{ss:cont_opa}. For line opacities the code takes a more sophisticated approach because line lists can contain tens to hundreds of thousands of lines and the majority of lines in a line list will not contribute significant opacity to any given spectral point (i.e., the spectral points far away from any given line). 
% and computing the contribution of each line to each requested spectral point is computationally inefficient, especially when the contribution of that line is negligible, i.e., to spectral points far away from the line.
Here, \stardis\ takes into account the line integrated opacity and the width of the line to select a portion of the frequency dependent opacity grid to apply the line to. The code then only calculates the Voigt profile and the opacity contribution of the line at relevant frequencies near the center of the line, scaling with line strength and width. 

% \subsubsection{Applied Line Opacity} \label{ss:alpha_line_at_nu} %I don't like this name

% With a complete description of the two parameters needed to determine the shape of any given line from Subsection \ref{ss:broadening}, (i.e., the Lorentzian full-width at half-max $\gamma$ and the Gaussian standard deviation $\sigma$), 

\subsubsection{Radiative Transfer} \label{ss:raytracing}

\stardis\ calculates the frequency-dependent optical depth $\tau(\nu)$ between each atmospheric depth point

\begin{equation}
    \tau(\nu) = n l \sigma(\nu) = \alpha(\nu) l
\end{equation}
where $l$ is the distance between two depth points, and $\alpha(\nu)$ is summed at each depth point detailed earlier in Section~\ref{ss:broadening} and Equation~\ref{eq:sum_opacity}. The summed opacities are logarithmically averaged between depth points. 

\stardis\ then solves the radiative transfer equation and computes a spectrum using the piecewise formal solver outlined in \citet{van_noort_multidimensional_2002}:

\begin{equation}
    I_2 = I_1e^{-\Delta\tau_{1,2}} + w_0S_2 + w_1 \frac{\partial S}{\partial \tau} \bigg|_2 + w_2 \frac{1}{2}\frac{\partial^2 S}{\partial \tau^2} \bigg|_2
\end{equation}
where $I$ is the intensity that is solved outwards from the inner boundary, $S$ is the source function, and the $w$ weight coefficients for the the expansion given by

\begin{gather*}
    w_0 = 1 - e^{-\Delta\tau_{1,2}}, \\
    w_1 = w_0 -\Delta\tau_{1,2}e^{-\Delta\tau_{1,2}},\\
    w_2 = 2w_1 - (\Delta\tau_{1,2})^2e^{-\Delta\tau_{1,2}}.
\end{gather*}
Using second-order finite difference approximations we have

\begin{equation}
    \frac{\partial S}{\partial\tau}\bigg|_2 = \frac{(S_2 - S_3)(\Delta\tau_{1,2}/\Delta\tau_{2,3}) - (S_2 - S_1)(\Delta\tau_{2,3}/\Delta\tau_{1,2})}{\Delta\tau_{1,2}+\Delta\tau_{2,3}}
\end{equation}

\begin{equation}
    \frac{1}{2}\frac{\partial^2S}{\partial\tau^2}\bigg|_2 = \frac{(S_3 - S_2)/\Delta\tau_{2,3} + (S_1 - S_2)/\Delta\tau_{1,2}}{\Delta\tau_{1,2}+\Delta\tau_{2,3}}
\end{equation}
The version of \stardis\ presented in this work uses the LTE approximation that the source function $S$ is purely the blackbody function.
% However, code is  the radiative transfer scheme implemented here is agnostic to the exact source function given and departures from LTE can be included at this step.
The radiative transfer equation is solved at each depth point for the given simulation (see Section~\ref{ss:atm_ingestion}).

\subsubsection{Spectral synthesis}\label{ss:spectrum}

\stardis\ uses a different spectral synthesis calculation depending on the geometry of the atmosphere. Currently, the options are ``plane parallel'' and ``spherical'', and are intended to be used with the corresponding model type. Solving the radiative transfer equation in spherical geometry mainly entails a modification to the paths that rays trace through the atmosphere, as well as casting rays from the far side of the star to accurately model edge rays \citep[see e.g.,][]{anusha_preconditioned_2009}. 

In either case, rays are traced for a number of angles specified by the user, and the weighted average is returned as astrophysical flux throughout the star, to capture the effects of rays emitting across the stellar surface. The rays are sampled and combined in a weighted average using Gauss-Legendre quadrature,

\begin{equation}
    F_\nu = \frac{\pi}{2}\sum_\theta^n w_\theta I_{\nu,\theta}
\end{equation}
The flux at the outermost point in the simulation (i.e. the surface of the star) is then the final calculated stellar flux.

\section{Comparison between several stellar spectral synthesis codes}\label{sec:code_comp}

\begin{figure*}[t] 
    \centering
    \includegraphics[width=.45\textwidth]{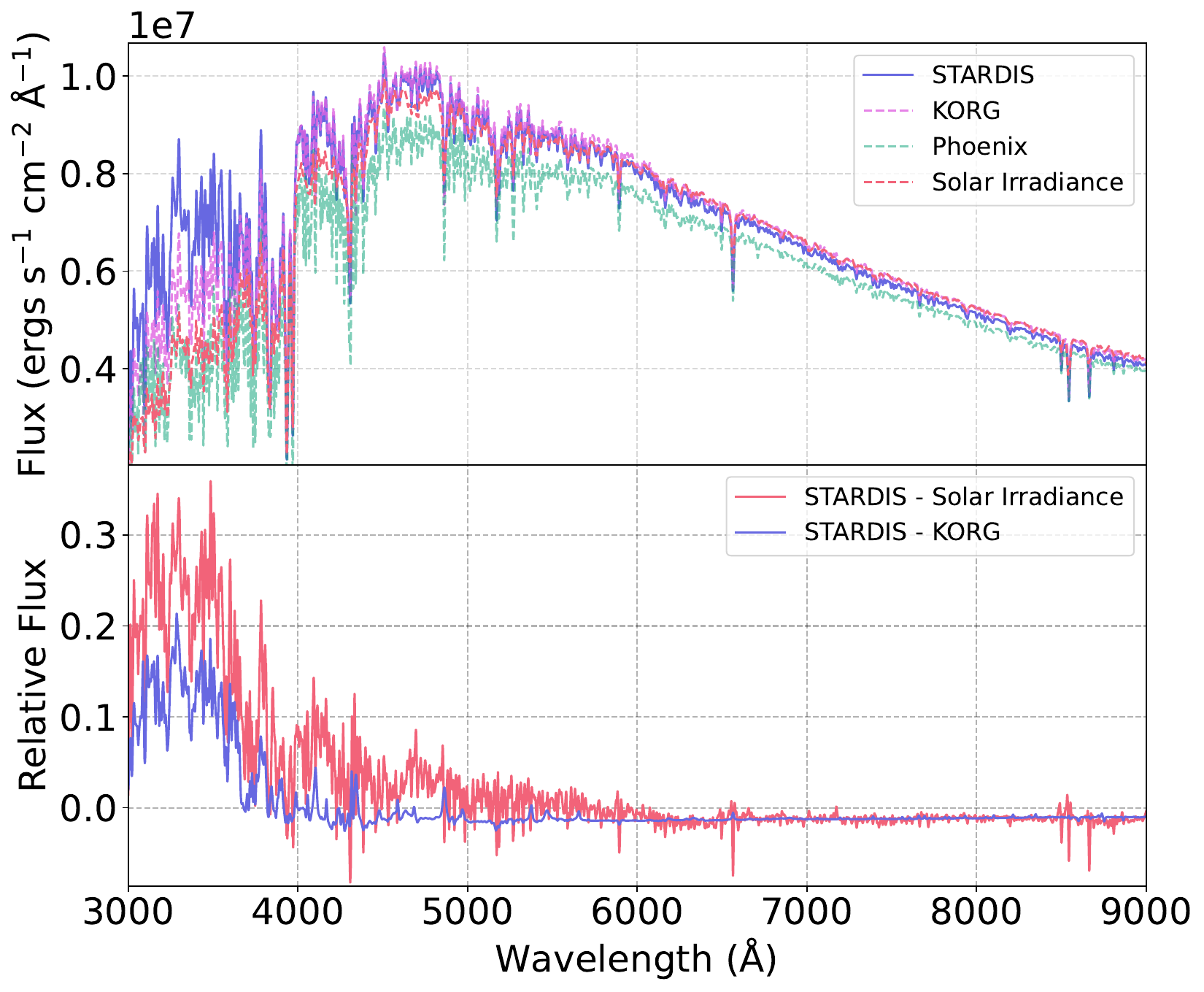}
    \includegraphics[width=.45\textwidth]{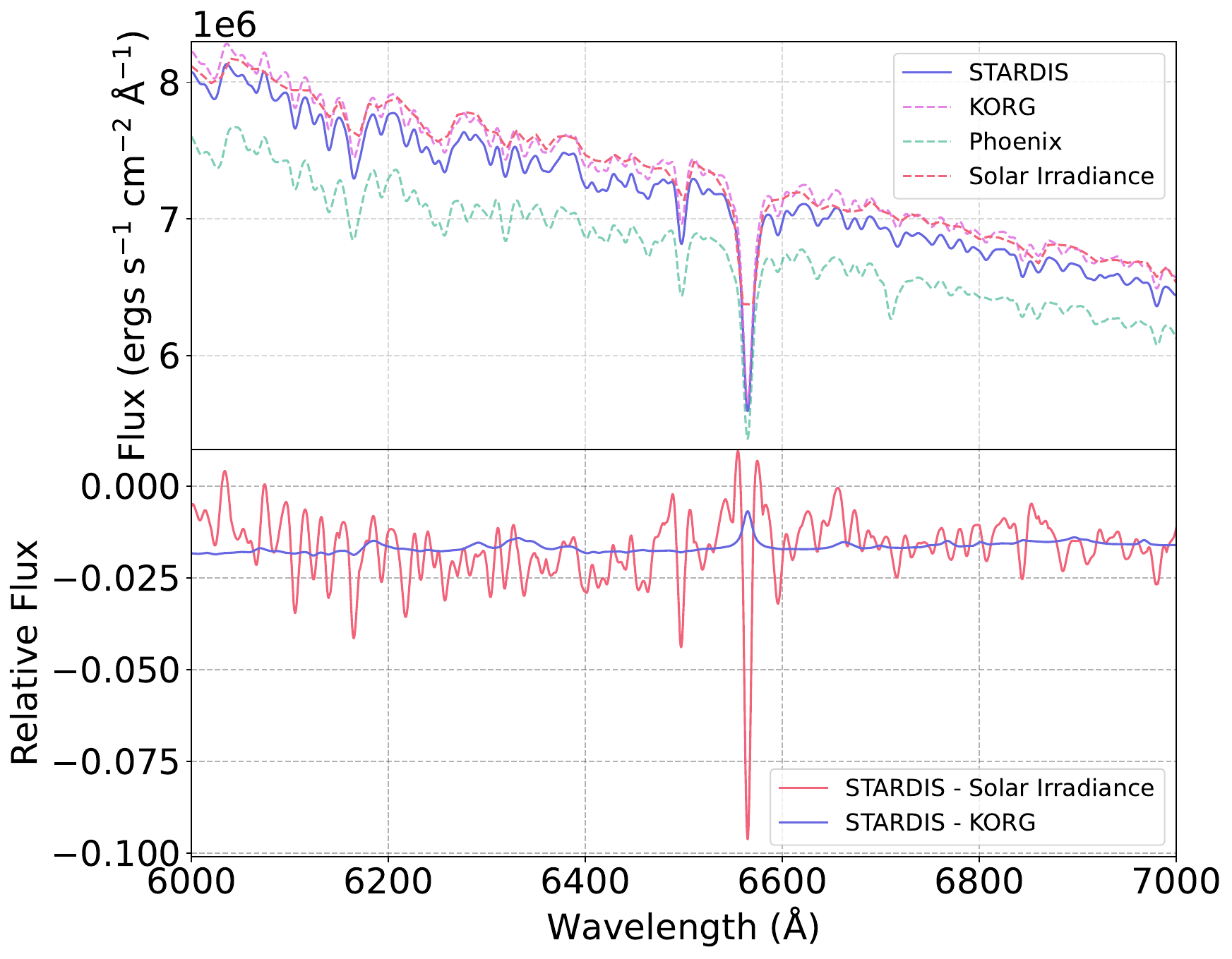}    
    \caption{A comparison to medium resolution (R$\ \approx\ $2000) spectral observations of the Sun. The very blue and UV spectra show poor agreement, often disagreeing above 10$\%$, but the raw flux of the spectrum beyond 5000~\AA\ shows strong agreement on typical instrumental resolutions.}
    \label{fig:solar_obs}
\end{figure*}

% ???? Maybe an introductory paragraph of what to expect: We have chosen the radiative transfer code Korg as the main source of comparison due to its ease of use and open-nature. In addition, we have performed select experiments with Phoenix [add to which ones you chose and why]???

We compare our code primarily to \textsc{korg} (v0.42.0) \citep{wheeler_korg_2023} evaluated over a variety of F, G, and K stars output by MARCS (last updated April 2018), and secondarily to \textsc{phoenix} (\citealt{hauschildt_fast_1992}, model presented in \citealt{husser_new_2013}) for the Sun specifically. We made this decision because \textsc{korg} is an open code which can generate spectra using the same inputs as \stardis, enabling the most direct comparison of the codes possible. In contrast, \textsc{phoenix} has readily available solar spectra, but the code is not publicly available so we cannot rerun the code to generate new spectra with new inputs (i.e., new atomic data and model atmospheres). We chose to use the same set of stars for comparison as used in \citet{wheeler_korg_2023}, with one modification, as shown in Table~\ref{tab:stars}. We also chose to use a microturbulence $\xi = 1.0$ km s$^{-1}$ for each model. As discussed in Section~\ref{ss:broadening}, we do not focus on the physical validity of microturbulence, but include and use it here to provide for as accurate a code comparison as possible. We synthesize spectra using with as similar metallicities to \textsc{korg} as possible, using \citet{asplund_chemical_2021} as a source for the base solar metallicity rather than raw metal number densities provided in the MARCS models themselves. Additionally, we synthesize spectra of $\alpha$ Centauri B instead of Arcturus to cover a large range of temperatures and surface gravities accurately modeled by \stardis. We emphasize that the exact choice of star here is less important than understanding the parameter spaces where the codes agree and disagree. 

\begin{table*}[ht] 
\centering
\caption{Stars Used For Spectral Code Comparison}
\begin{tabular}{l|l|l|l|l} \label{tab:stars}
                    & $T_\textrm{eff}$ (K) & log$g$ & $[\frac{\textrm{M}}{\textrm{H}}]$ & $\xi$ (km s$^{-1}$)\\ \hline
The Sun & 5777 & 4.44 & 0.0 & 1.0 \\ 
$\alpha$ Centauri B & 5250 & 4.0 & 0.25 & 1.0   \\ 
HD122563 & 4500 & 1.5 & -2.5 & 1.0 \\
HD499330 & 6250 & 4.0 & -0.5 & 1.0

\end{tabular}
\tablecomments{The four stars shown for code comparison and validation, and the astrophysical parameters of the nearest models available. MARCS models with the parameters listed were used to generate the spectra shown in Figures~\ref{fig:sol_comp} -- \ref{fig:6250_comp}.}
\end{table*}

% We show a collection of synthesized spectra of four different stars across F, G, and K types [??? add why you chose those - saying they are extensively tested in the korg paper is good enough] (see Table~X) synthesized comparing to spectra synthesized by  \textsc{korg} \citep[see][]{wheeler_korg_2023} in Figures~\ref{fig:sol_comp}-\ref{fig:6250_comp}.

For our direct comparisons we have chosen four characteristic wavelength regions with fluxes sampled every 0.1~\AA\ in each spectrum. We chose the 6563~\AA\ H${\alpha}$ line as one of the strongest stellar lines to validate our broadening prescription and understand how different broadening prescriptions affect the Voigt profile. We show two lines of the \ion{Ca}{2} triplet at 8500~\AA\ to investigate our spectral synthesis in the near infrared regime. For each of these windows, we chose to normalize each spectrum by removing the continuum. We find the continuum of the \stardis\ and \textsc{korg} spectra by taking full simulation outputs generated without spectral lines, and dividing by fluxes on a per-wavelength basis. We also show the \ion{Ca}{2} K line as a characteristic near ultraviolet test, but for this window we choose to show absolute rather than normalized fluxes to highlight the disagreement between the codes in ultraviolet continua, rather than line shape and strength. Finally, we show a \ion{Mg}{1} feature near 5185~\AA\ to give us an understanding of the green optical regime as well as the diatomic C$_2$ band between 5160--5165~\AA\ that appears in stars cold and metal rich enough to permit molecular formation. 

We see a general trend that the strongest lines in each spectrum for each star spectrum agree remarkably well, with differences appearing most significantly in the UV continuum. 

% These windows focus on the 6563~\AA\ H-$\alpha\ $line (??? say why -- I think to test broadening parameters???, the Ca triplet around 8600~\AA, the CA K line at 3970~\AA, and the strong Mg I line at 5185~\AA. This final plot also shows a strong diatomic molecular C feature from around 5160 to 5168~\AA, in atmospheres cold enough for the molecule to form.   To illustrate this, Figure~\ref{fig:cntm_comp} shows a broad solar comparison with all synthetic spectra convolved to an instrumental resolution R = 2000 alongside an extraterrestrial solar spectrum from \citet{gueymard_revised_2018}. All comparisons between \stardis\ and \textsc{korg} shown were generated using the same MARCS atmospheres and the same VALD line list. 

\begin{figure*}[t] 
    \centering
    \includegraphics[width=.45\textwidth]{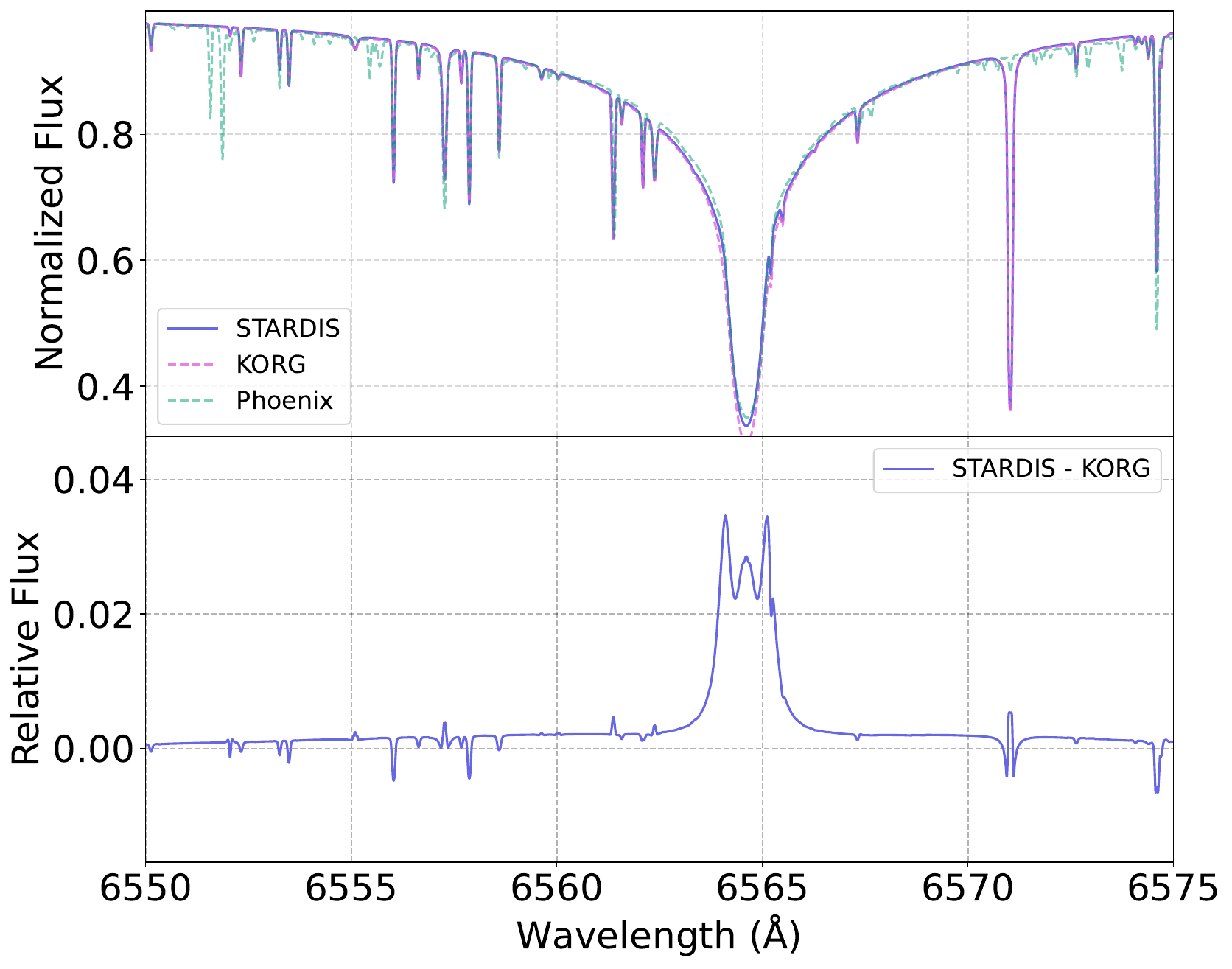}
    \includegraphics[width=.45\textwidth]{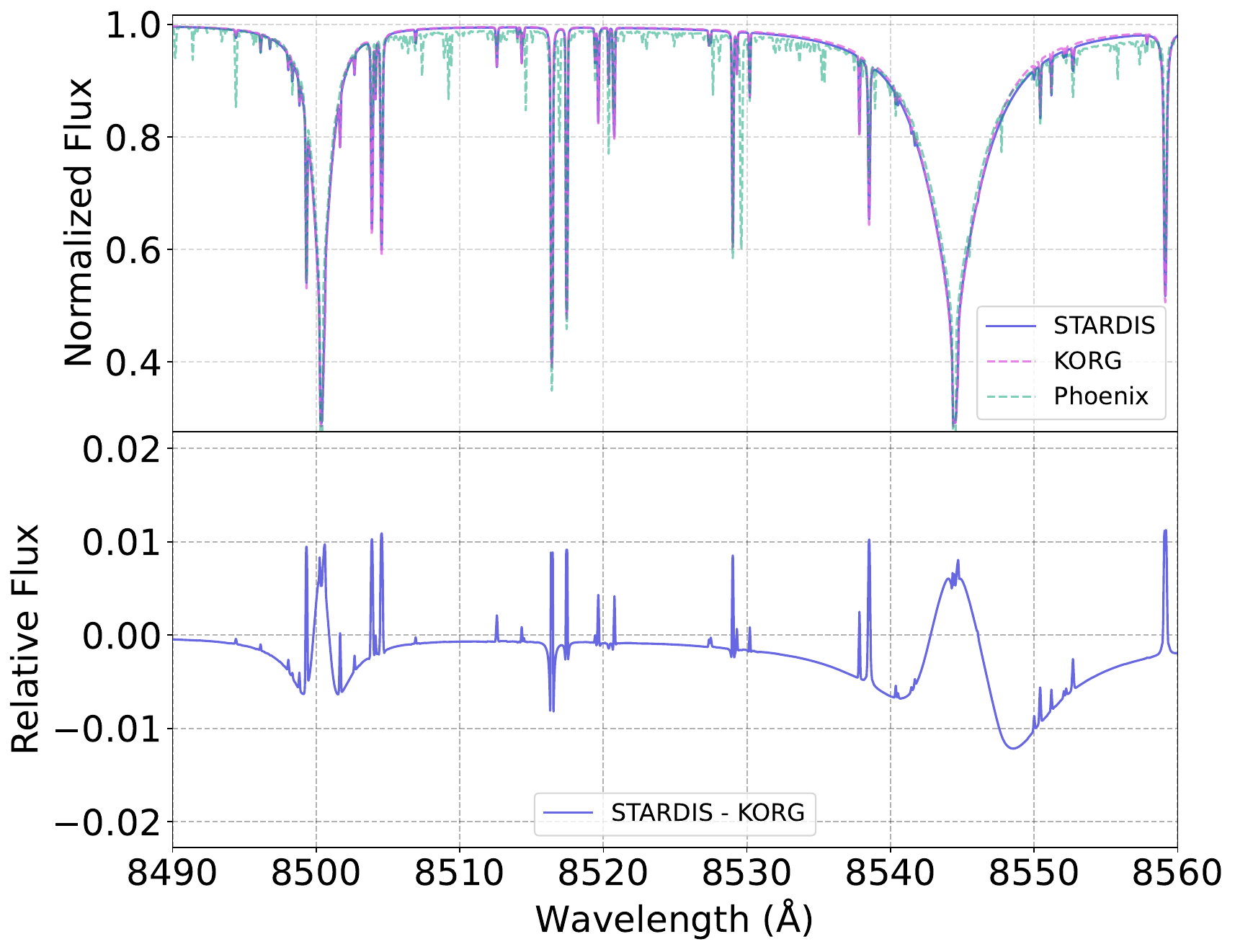}
    \includegraphics[width=.45\textwidth]{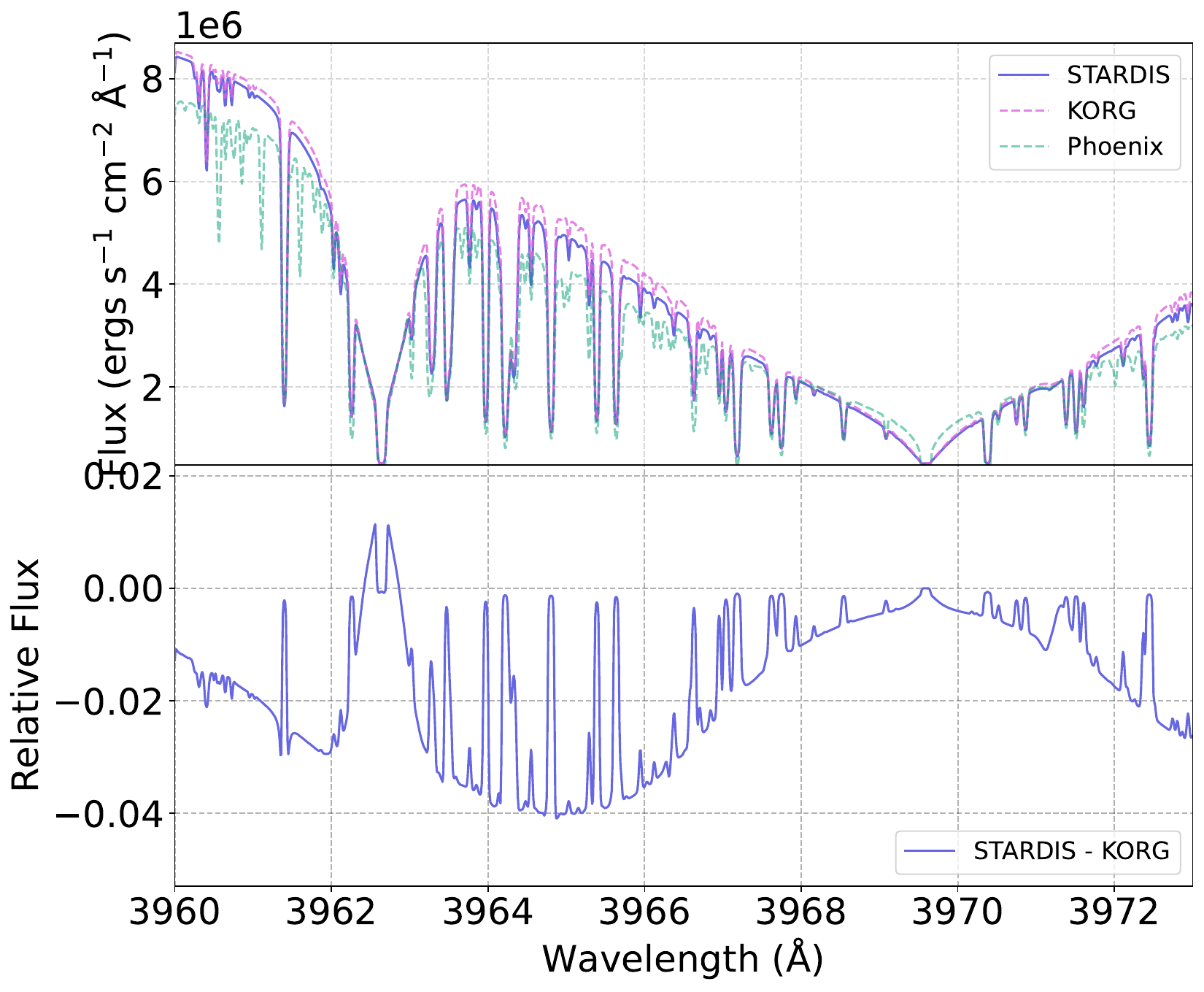}
    \includegraphics[width=.45\textwidth]{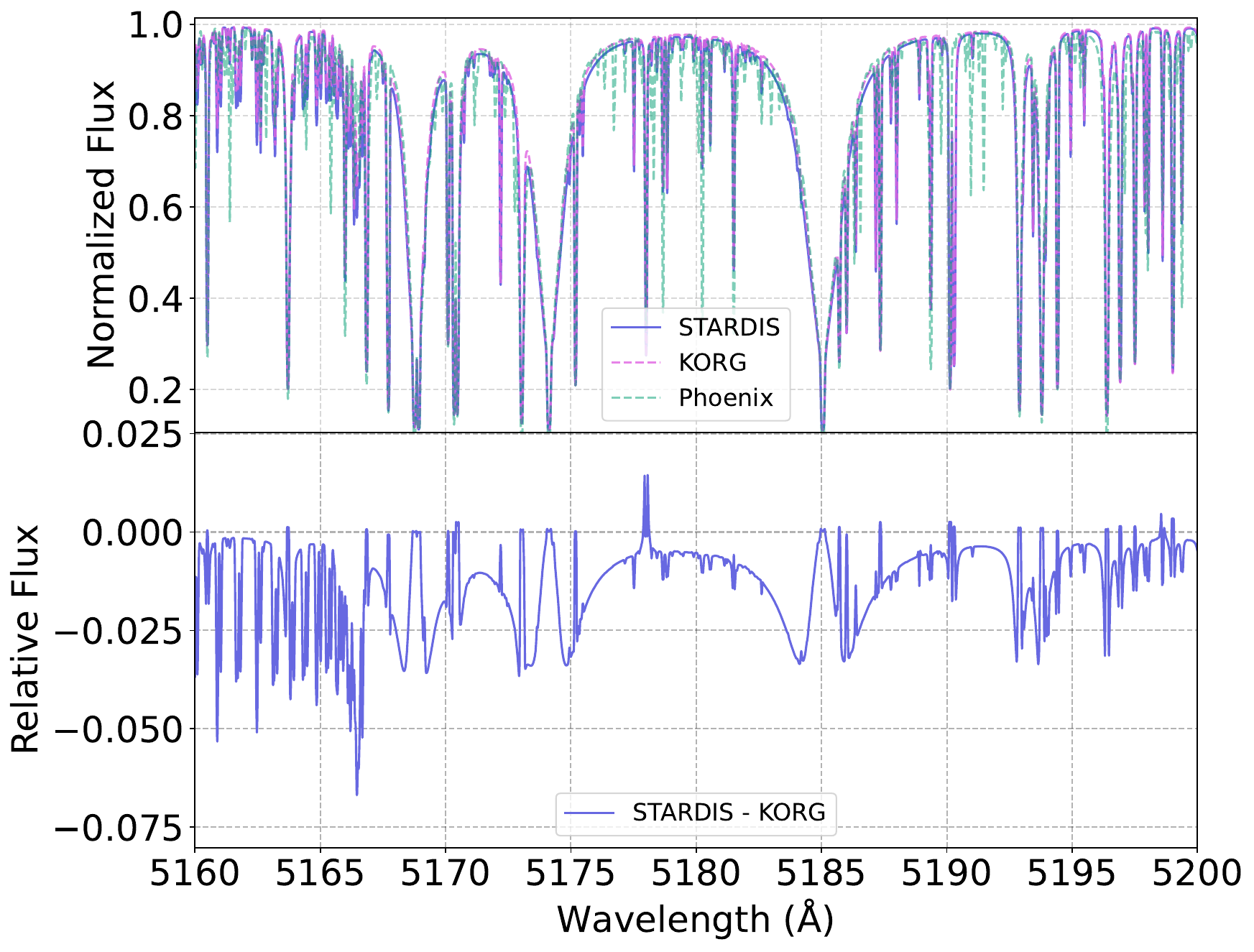}    
    \caption{A spectral comparison between \stardis\ and \textsc{korg} for a solar analog, with stellar parameters $T_\textrm{eff} = 5777$ K, $\log g = 4.44$, $[\frac{\textrm{M}}{\textrm{H}}] = 0.0$, $\xi = 1.0$\ km\,s$^{-1}$. The top left shows a focus on the H$\alpha$ line. The top right shows the \ion{Ca}{2} triplet. The bottom left shows the \ion{Ca}{2} K line. The bottom right shows a strong \ion{Mg}{1} line. All spectra here are continuum normalized except the \ion{Ca}{2} K line to show discrepancies in UV continuum flux.}
    \label{fig:sol_comp}
\end{figure*}

\begin{figure*}[t] 
    \centering
    \includegraphics[width=.42\textwidth]{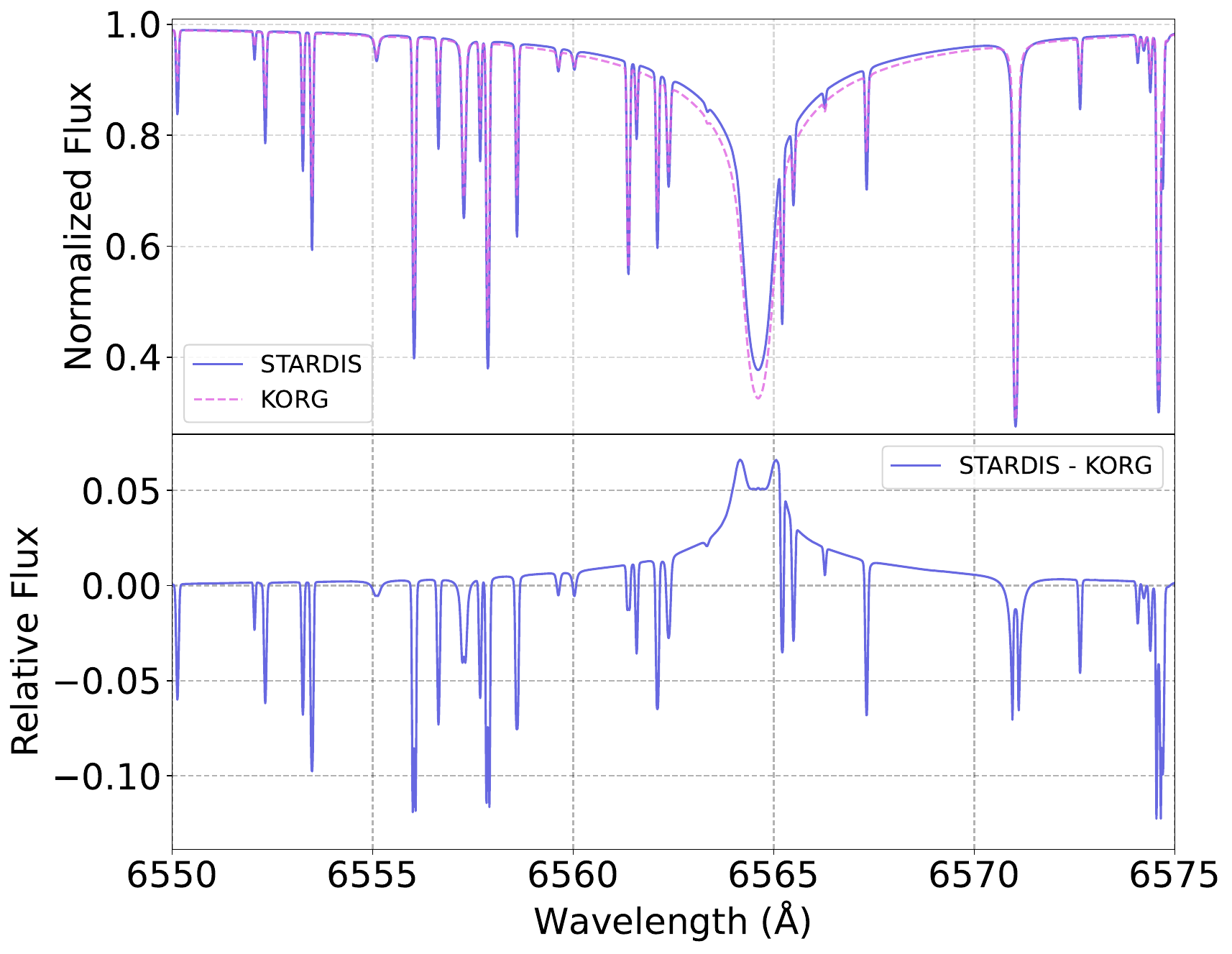}
    \includegraphics[width=.42\textwidth]{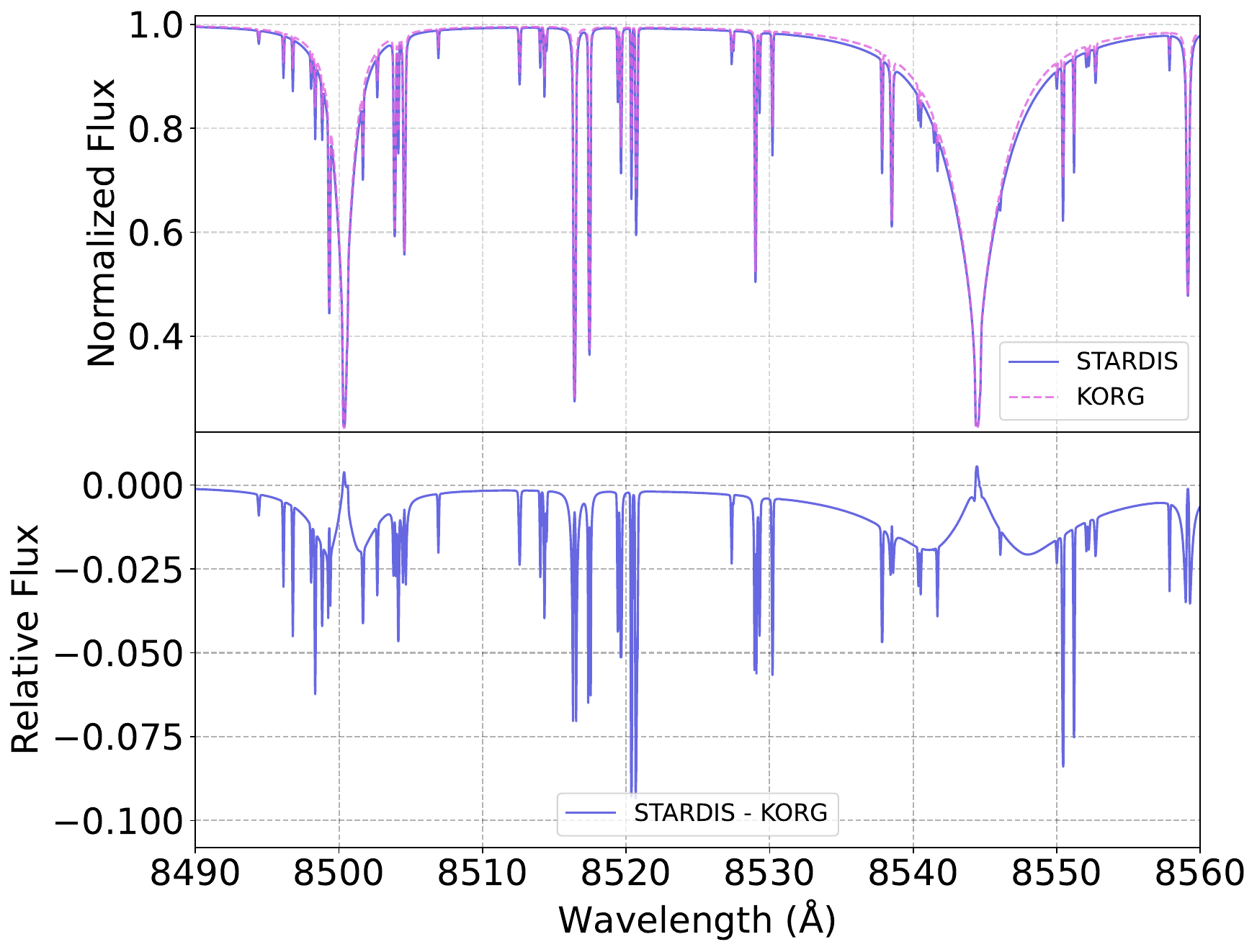}
    \includegraphics[width=.42\textwidth]{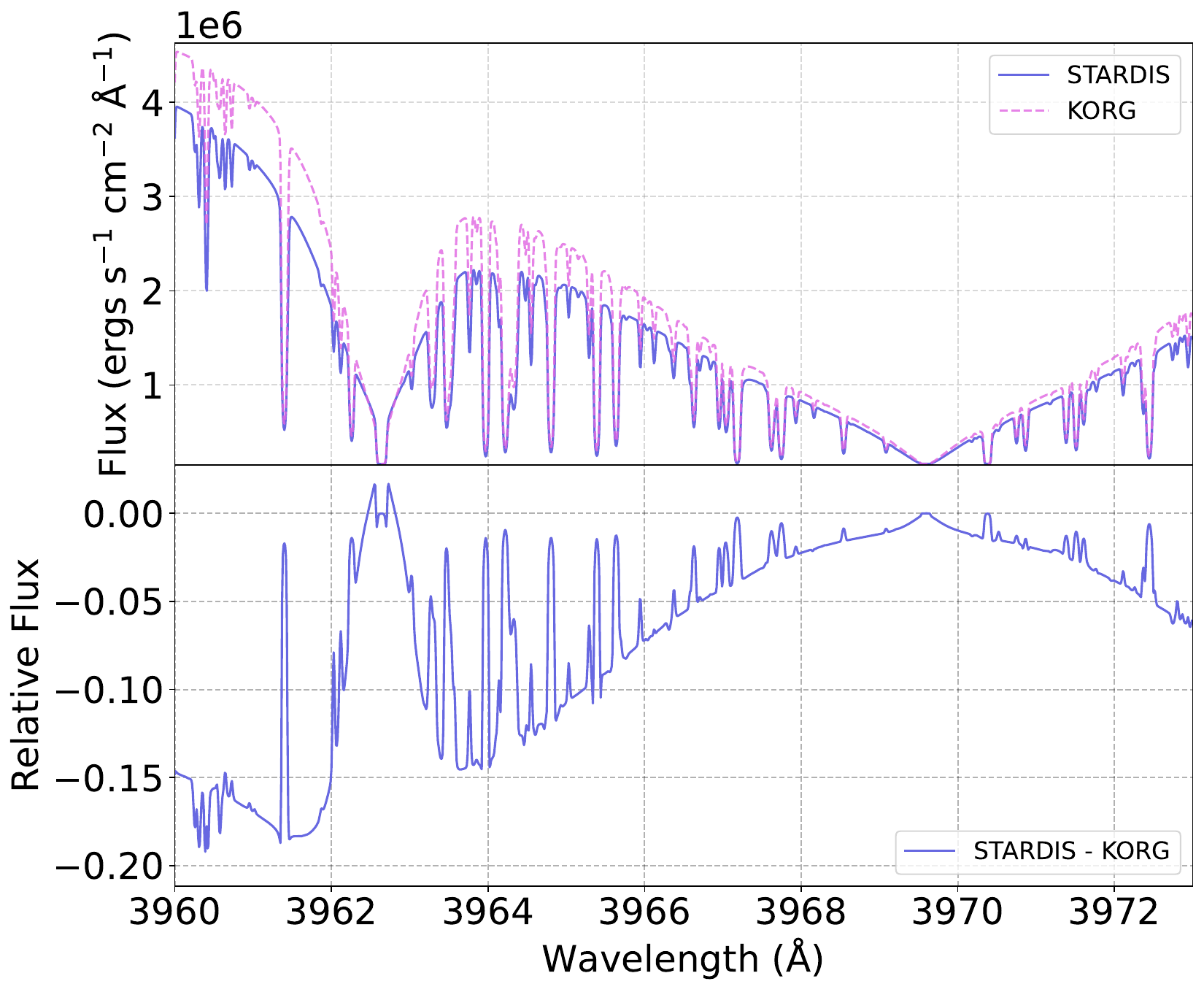}
    \includegraphics[width=.42\textwidth]{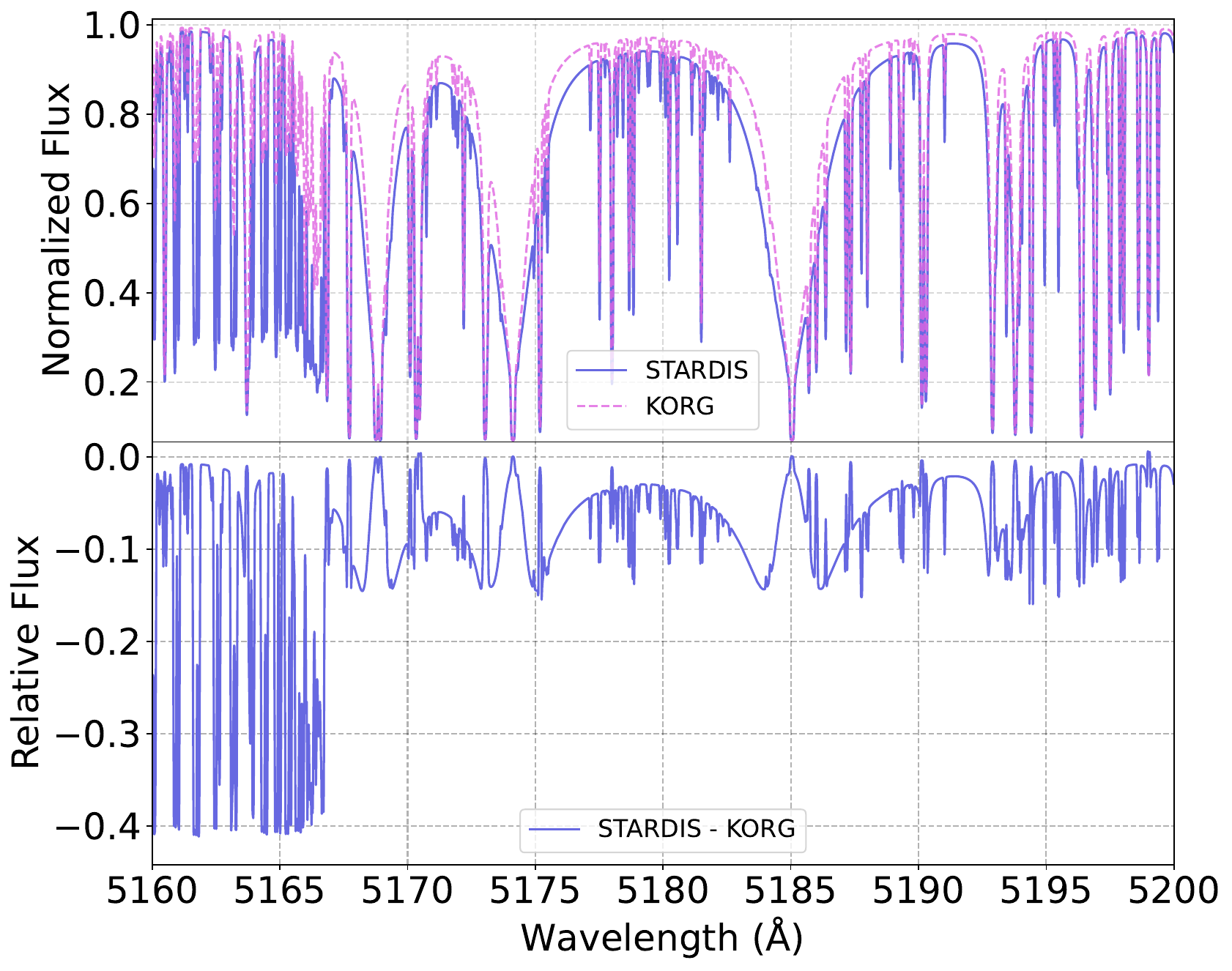}    
    \caption{Similar to Figure~\ref{fig:sol_comp}, but for an $\alpha$ Centauri B like star with stellar parameters $T_\textrm{eff} = 5250$~K, $\log g = 4.0$, $[\frac{\textrm{M}}{\textrm{H}}] = 0.25$, $\xi = 1.0$~km~s$^{-1}$. The most metal-rich star in the sample. }
    \label{fig:5250_comp}
\end{figure*}

\begin{figure*}[t] 
    \centering
    \includegraphics[width=.45\textwidth]{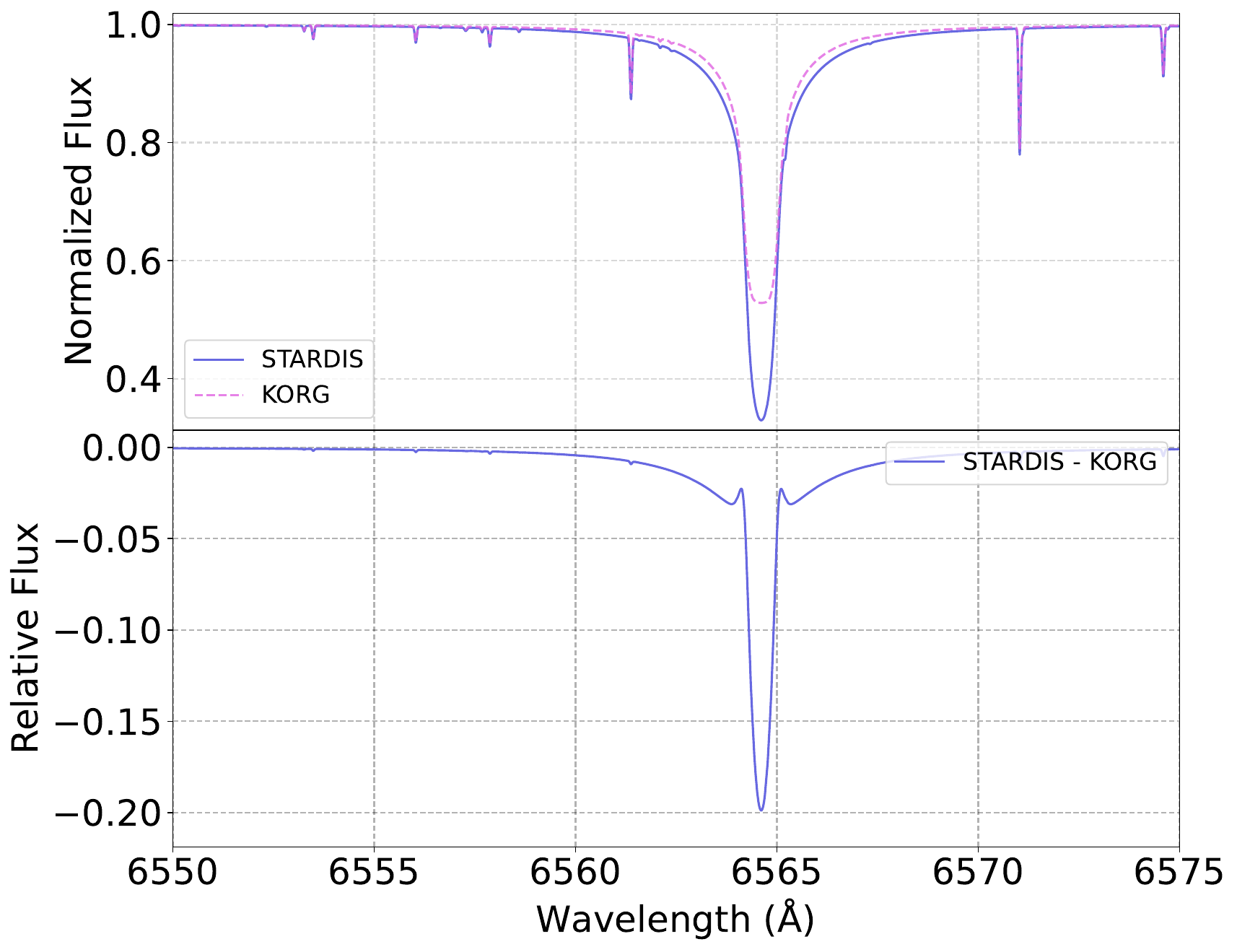}
    \includegraphics[width=.45\textwidth]{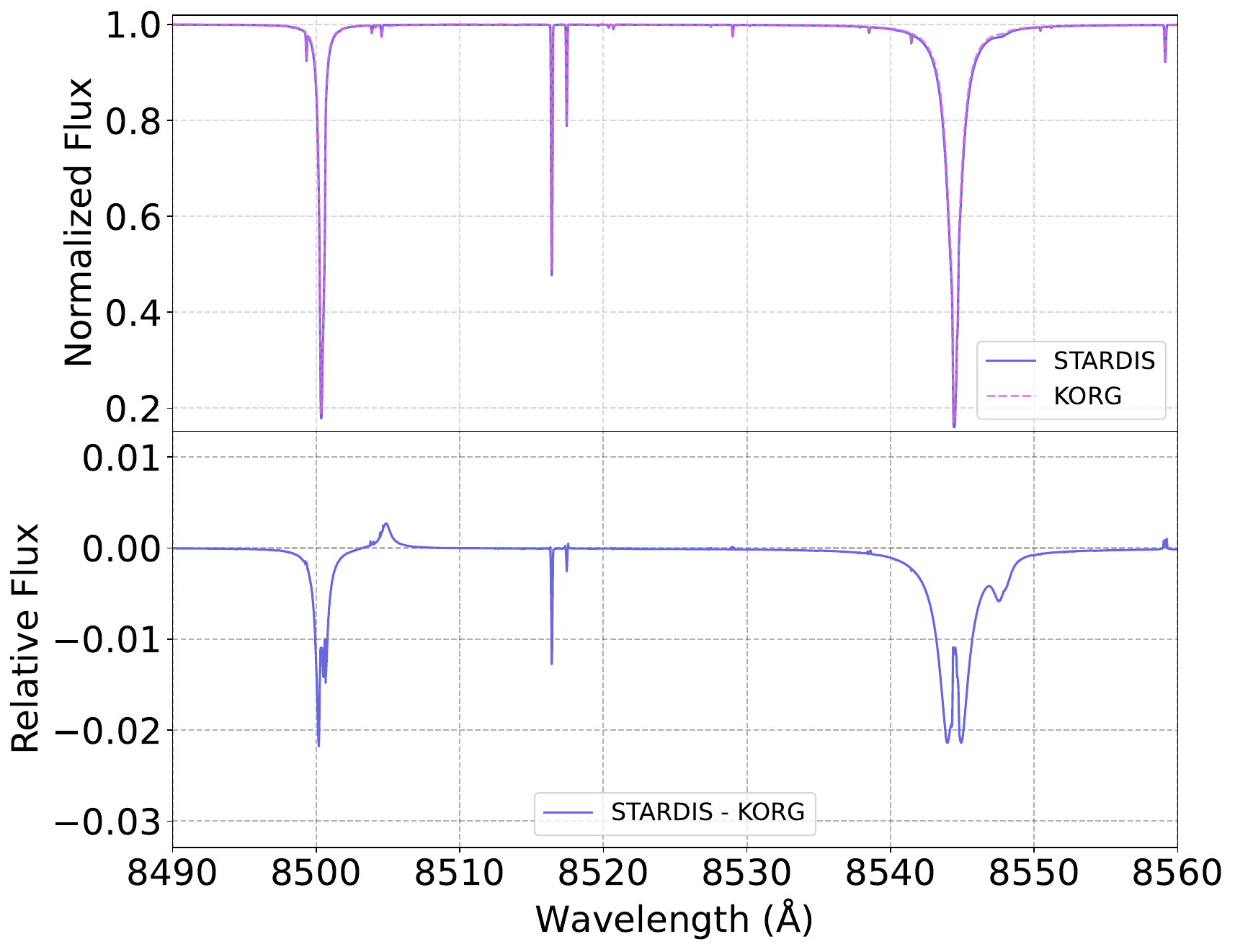}
    \includegraphics[width=.45\textwidth]{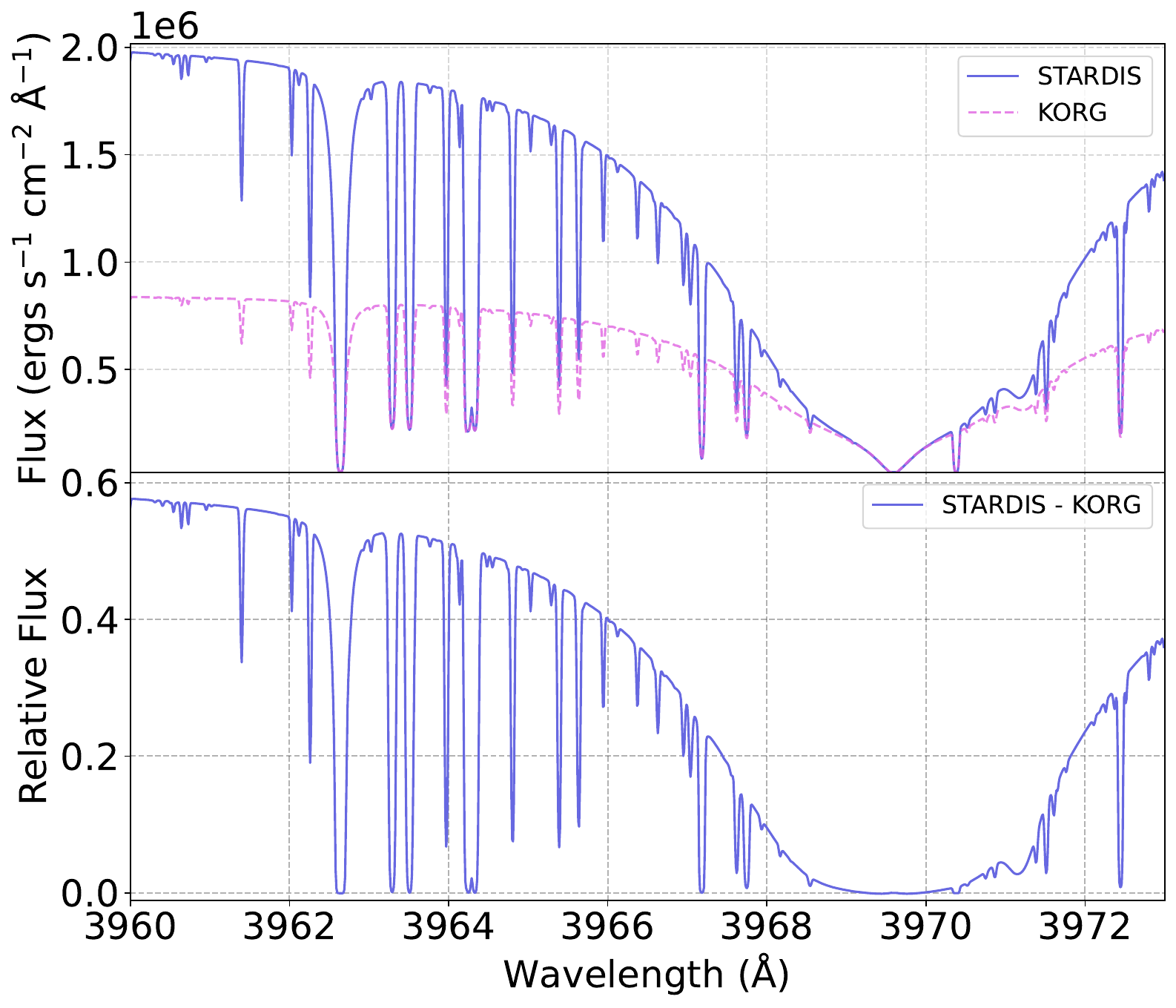}
    \includegraphics[width=.45\textwidth]{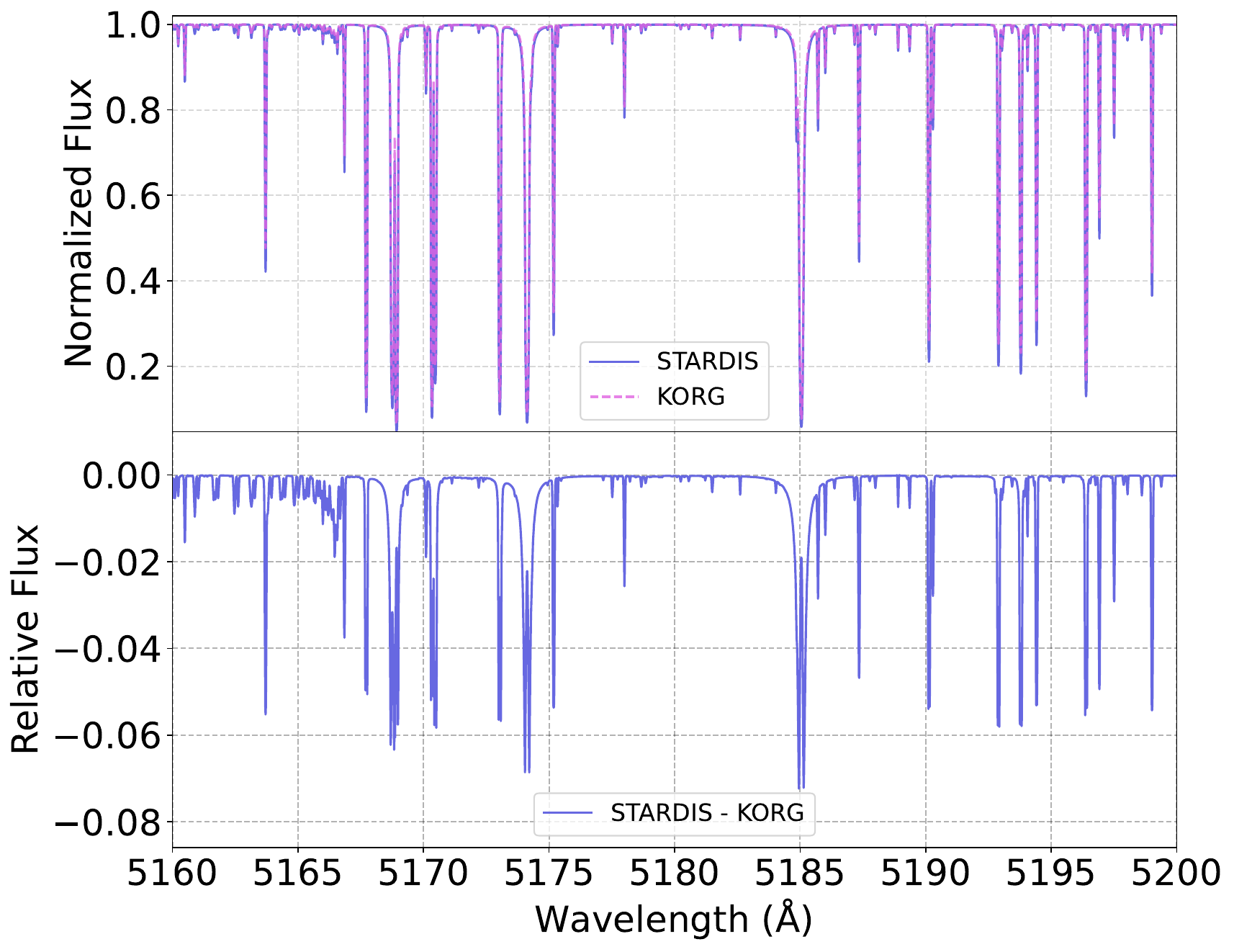}    
    \caption{Similar to Figure~\ref{fig:sol_comp}, but for a K star similar to HD122563, an extremely metal-poor red giant, and the coolest star in our sample. This star has stellar parameters $T_\textrm{eff} = 4500$~K, $\log g = 1.5$, $[\frac{\textrm{M}}{\textrm{H}}] = -2.5$, $\xi = 1.0$~km~s$^{-1}$.}
    \label{fig:4500_comp}
\end{figure*}

\begin{figure*}[t] 
    \centering
    \includegraphics[width=.45\textwidth]{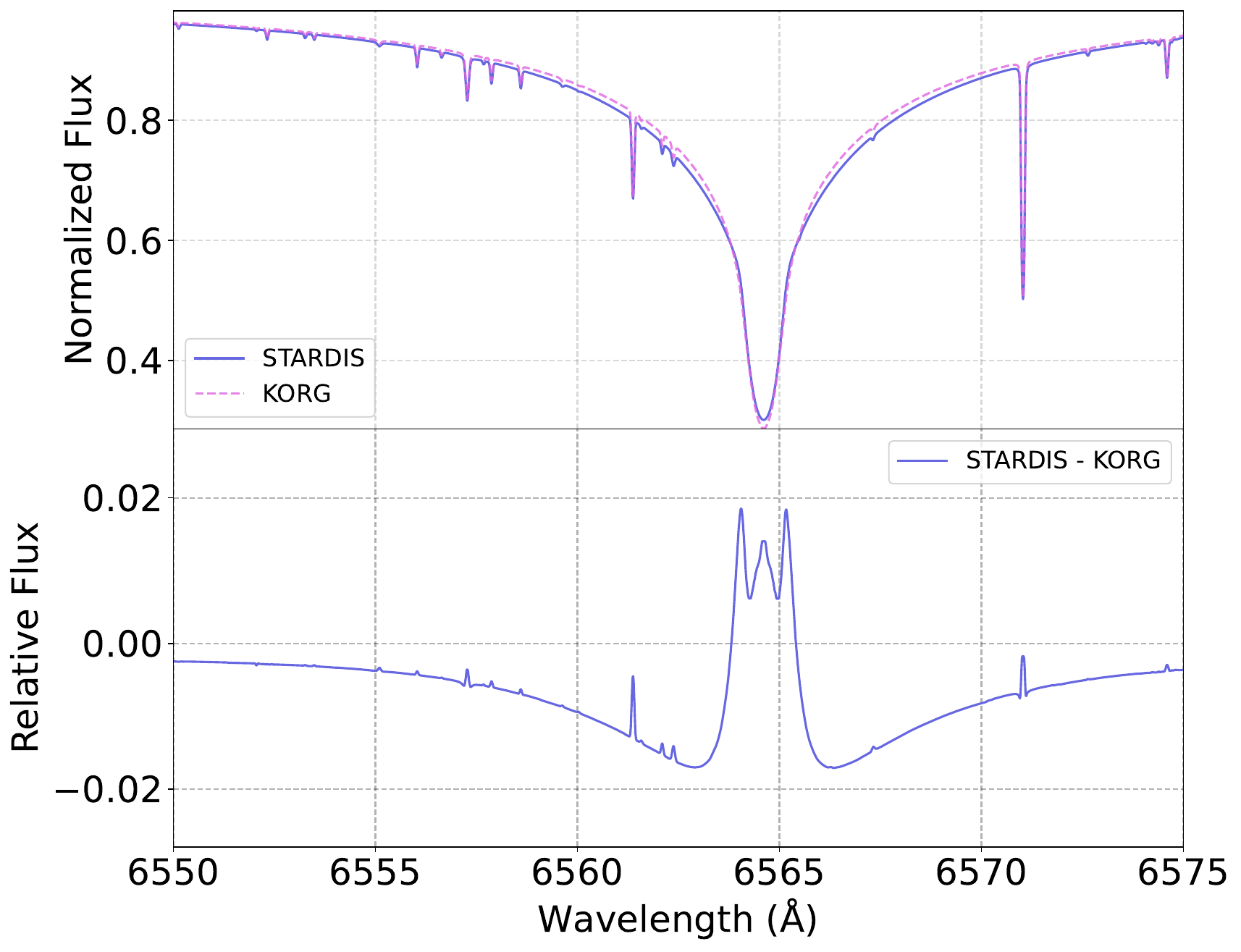}
    \includegraphics[width=.45\textwidth]{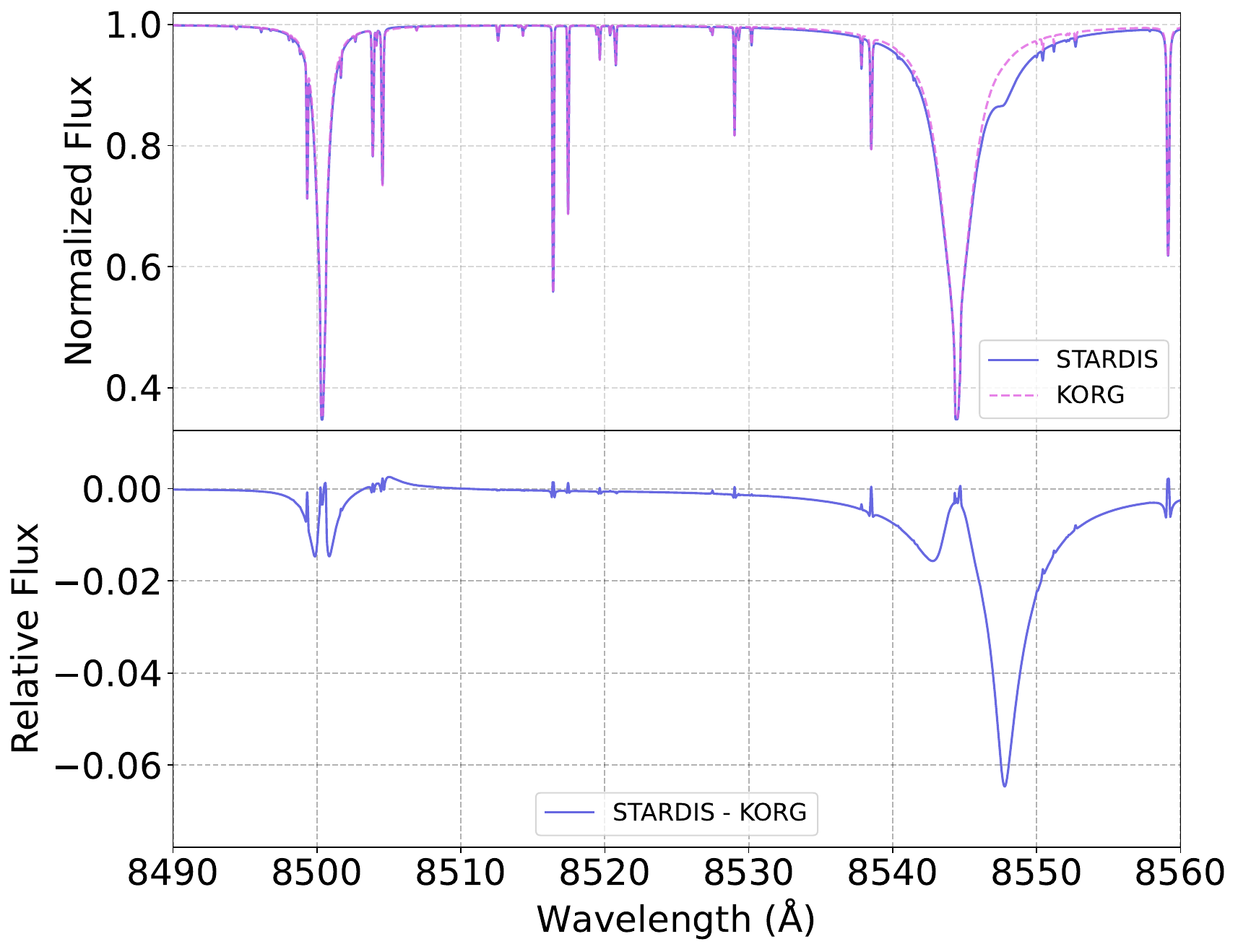}
    \includegraphics[width=.45\textwidth]{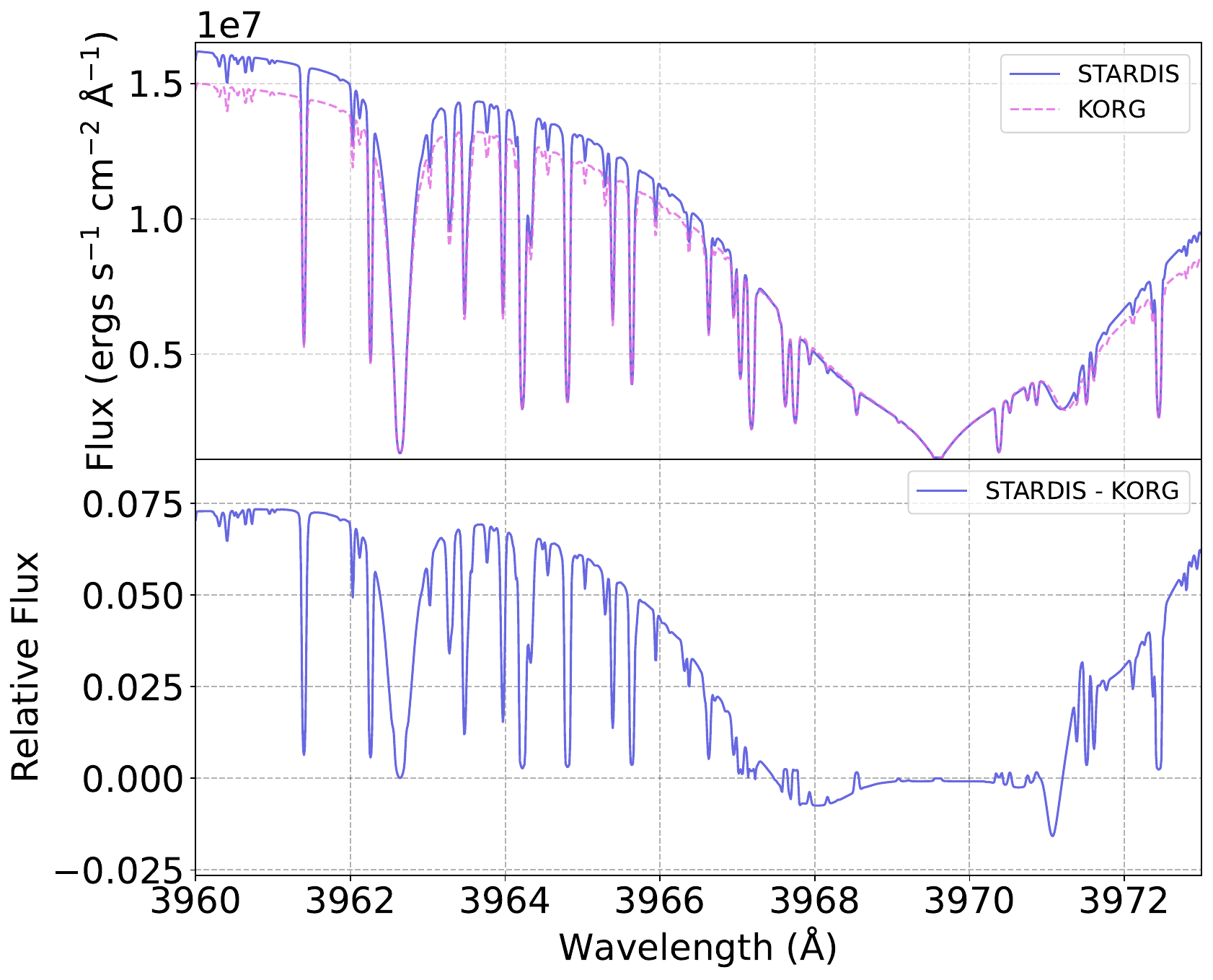}
    \includegraphics[width=.45\textwidth]{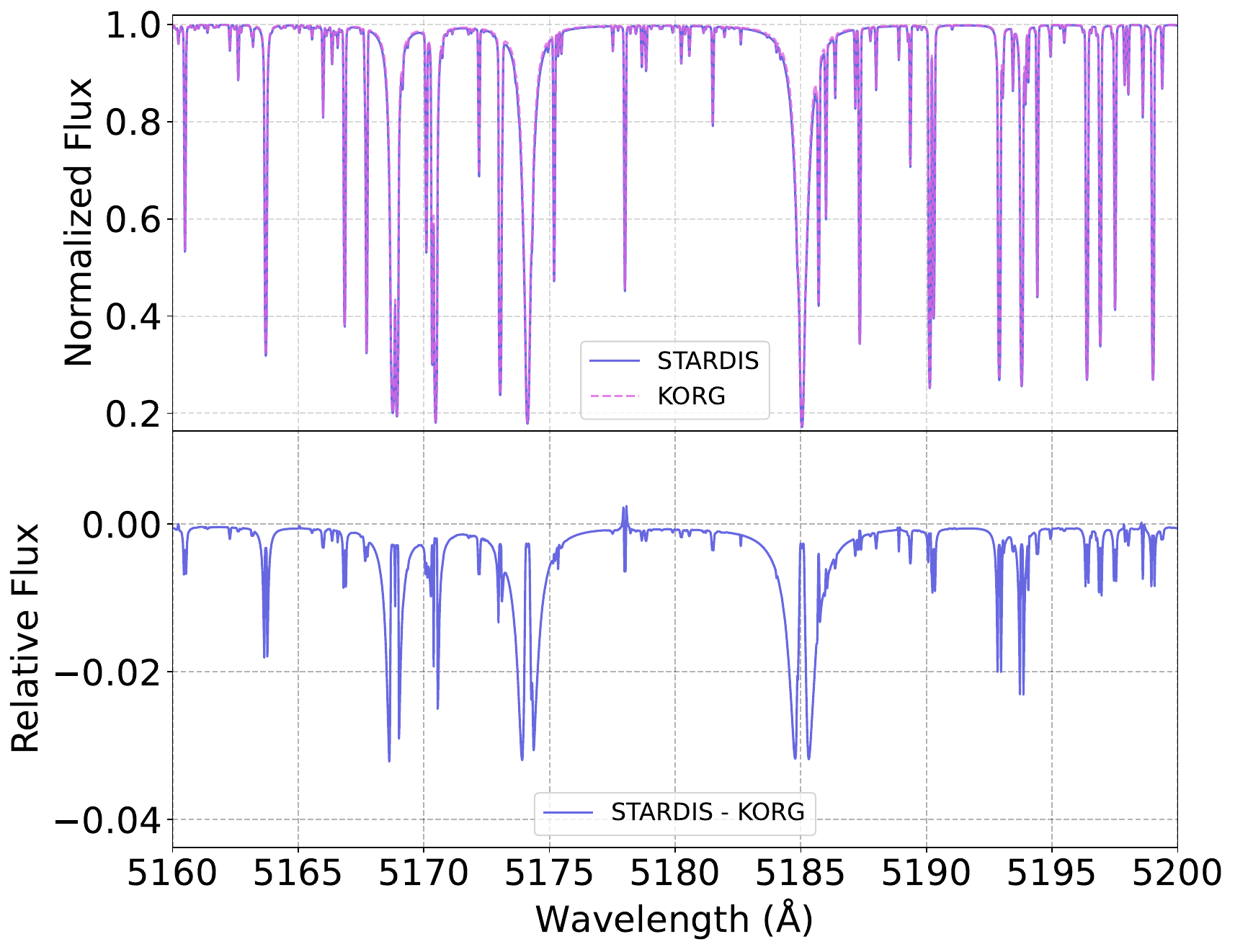}    
    \caption{Similar to Figure~\ref{fig:sol_comp}, but for a metal poor F dwarf star similar to HD499330, the hottest star in our sample, with stellar parameters $T_\textrm{eff} = 6250$~K, log$\, g = 4.0$, $[\frac{\textrm{M}}{\textrm{H}}] = -0.5$, $\xi = 1.0$~km~s$^{-1}$.}
    \label{fig:6250_comp}
\end{figure*}

\subsection{The Sun}\label{ss:sun}

Our spectral synthesis of the Sun shows strong agreement  across most of the spectrum, with differences appearing primarily in a few missing lines from the \stardis\ spectrum. We begin by showing a broad solar comparison in Figure~\ref{fig:solar_obs}. Here, all spectra are convolved to an instrumental resolving power of $R = 2000$ alongside an extraterrestrial solar spectrum from \citet{gueymard_revised_2018}. The left shows the entire synthesized spectrum, while the right focuses on a 1000 \AA range to allow for closer investigation of individual line features and a typical optical comparison. \stardis\ and \textsc{korg} agree on raw fluxes at a level of 2\% or better everywhere redder than ultraviolet wavelengths. At the bluer wavelengths, line blanketing of large numbers of unresolved lines becomes important which have historically been difficult to model, though potentially within computational reach of current hardware. Furthermore exact choices of continuum calibration become significant at this precision which is beyond the scope of this work. We choose to show this wavelength region as a checkpoint for the current state of ultraviolet spectral synthesis, but focus on the redder wavelengths for the majority of our analysis and validation. However, we do note that \textsc{korg} includes several sources of continuum flux that become important in the UV regime which have not yet been implemented in \stardis\ and likely explains a significant portion of the disagreement between the codes here. 

We show a detailed comparison of the four wavelength windows discussed earlier in Figure~\ref{fig:sol_comp}. These figures also show a \textsc{phoenix} spectrum. However, the \textsc{phoenix} spectra shown were generated with different atomic data and a different metallicity profile (i.e., not from \citealt{asplund_chemical_2021}). We show \textsc{phoenix} spectra for qualitative comparison of large, shared lines between the three codes, but stress that the presence or lack of lines between \textsc{phoenix} and \stardis\ or \textsc{korg} is primarily due to the different atomic data and composition. This is also why we choose not to show residuals to \textsc{phoenix}, as those residuals would be dominated by these systematic differences. 

In comparing \stardis\ to \textsc{korg}, the similarity in the H$\alpha$ line to better than 4\% between the two codes is particularly noteworthy. The \stardis\ broadening prescription for H is dramatically different from that of \textsc{korg}, as \textsc{korg} follows methodologies laid out in \citet{barklem_hydrogen_2003} and described in \citet{wheeler_korg_2023}, while \stardis\ treats H lines similarly to every other line present in the simulation using VALD broadening parameters. 
%This is particularly interesting, as it disagrees with the idea that H needs special treatment. 

The \ion{Ca}{2} triplet window shows strong agreement to better than 1\% over all wavelengths, with individual frequencies agreeing more closely than observations could distinguish. This is a broad trend that persists for each of the comparison stars, but highlights the similarity of \stardis and \textsc{korg} for the majority of lines at redder wavelengths. 

The \ion{Mg}{1} window shows some of the strongest disagreement as a result of the inclusion of the C$_2$ band from 5160--5165~\AA. Codes often disagree on due to differences in the C$_2$ dissociation energy incorporated into the code. To generate these spectra, \stardis\ implicitly used a dissociation energy from \citet{barklem_partition_2016} incorporated into the molecular equilibrium constants discussed in Section \ref{ss:molecules}, which have similarly been incorporated into \textsc{korg}. The lines themselves disagree on the level of about 3\%, which is likely due to the different molecular solutions employed by the two codes.

It warrants discussion that the solar spectral synthesis performed by \stardis\ in this work overall is remarkably similar to the spectrum generated by \textsc{korg}. This is true in spite of the largely different implementations the two codes have in terms of specific broadening treatments and chemical number density calculations. We emphasize that the solar spectrum is a natural calibration target in the development of any spectral synthesis code, \stardis\ notwithstanding. That is, \stardis\ was not developed to closely match the \textsc{korg} solar spectrum, but likely both codes were developed with the Sun as a target. 

\subsection{$\alpha$ Centauri B}\label{ss:centauri}

$\alpha$ Centauri B, shown in Figure~\ref{fig:5250_comp}, is the most metal rich star we test in this work. The star is also cold enough to allow for significant amounts of molecular formation, which is exacerbated by the high metallicity. 
Consequently, the most noteworthy point of disagreement in these spectra is the C$_2$ feature around 5165~\AA. Particularly for metal rich or cold stars, current \stardis\ plasma calculations over-predict molecular formation when a significant fraction of atomic components would form multiple molecules. Currently, \stardis\ does not incorporate molecular formation into the full equilibrium balance solvers, and reactants can be counted multiple times instead of being taken out of the plasma when reactants form a molecule. If there are multiple molecules that share a common reactant, those molecules may be overestimated, but this effect diminishes at temperatures above 4250K with solar or metal poor compositions. 

\stardis\ predicts marginally stronger features throughout the spectrum for many of the narrow lines that appear. This suggests a systematic difference somewhere in the process of how \stardis\ and \textsc{korg} calculate the optical depth of the lines. This could be due to a somewhat higher number density of the elements responsible for the lines, but may also be due to a difference in the radiative transfer solution that differently weights certain parts of the atmosphere. 

The non-normalized \ion{Ca}{2} K window shows an interesting feature that appears in other comparison stars as well, where the strength of lines themselves is remarkably consistent between the two codes, while the width of the Lorentzian wings in the large line itself somewhat disagrees. This suggests that \stardis\ may disagree with \textsc{korg} on the specific broadening parameter that is dominant for this line and in this star. From Section \ref{ss:broadening}, we note that Stark broadening scales with electron number density, and the higher metallicity of this star may lead to a significantly higher electron density that results in that specific broadening term becoming more dominant for this star than others, and may explain the discrepant profiles.

% This is likely due to the disconnected nature of current \stardis\ molecular formation, which can struggle with super-solar metallicities, or effective temperatures below around 4250 K. 
%COME BACK HERE AND DOUBLE CHECK

\subsection{HD122563}\label{ss:4500K}

Figure~\ref{fig:4500_comp} shows a detailed comparison of our chosen wavelength regions for an extremely metal-poor HD122563 like giant star. HD122563 is the coldest star we investigate in this work. For this reason, it provides a useful test case in understanding how the strength and profiles of lines in the spectrum scale to lower temperatures in more extended atmospheres, as well as a test of how \stardis\ solves the radiative transfer equation in spherical geometry. Non-extended stellar atmospheres are largely solved with the plane-parallel assumption, which is valid when the distance between depth points is small compared to the curvature of the atmosphere, and the assumption must be relaxed for giant atmospheres. 

Here, the predicted H$\alpha$ strength disagrees significantly, with \stardis\ predicting a stronger absorption feature by roughly 20\% than \textsc{korg}. However, as \citet{wheeler_korg_2023} details, \textsc{korg} predicts a weaker H$\alpha$ line than \textsc{turbospectrum}, \textsc{SME}, or the observed spectrum of the star itself. 

We more generally see that \stardis\ again predicts marginally stronger lines than \textsc{korg}, which is most pronounced in the \ion{Mg}{1} window. The disagreement is significantly lessened than compared to the residuals seen in $\alpha$ Centauri B, which lends further strength to the argument that \stardis\ and \textsc{korg} might disagree on how lines scale to electron dense environments. We emphasize that both \textsc{korg} and \stardis\ read atmospheric temperatures and densities directly from model files, so the discrepancies are not due to these quantities directly.

% We more generally see that \stardis\ predicts somewhat stronger, broader lines than \textsc{korg} in this parameter space, especially in the Lorentzian wings. As \stardis\ has been set to ingest VALD broadening constants for these comparisons, the same as \textsc{korg}, and both codes read their atmospheric temperatures directly from the same MARCS model files, the most likely source of the discrepancy is due to discrepant number densities of broadening sources and line opacities themselves. For instance, from Section \ref{ss:broadening}, we might expect that a broader Lorentzian could be due to a somewhat higher electron or \ion{H}{1} density in the relevant portion of the atmosphere. Alternatively, the discrepancy could be due to somewhat different radiative transfer solvers themselves, subtly weighting parts of the atmosphere or paths through the atmosphere with different strengths. 

\subsection{HD499330}\label{ss:6250K}

Figure~\ref{fig:6250_comp} focuses on a comparison for the hottest star tested and compared in this work. Hotter stars show weaker metal absorption lines due to generally more ionized atoms in the atmosphere. As the hottest star shown, it provides another test case to examine how line profiles scale in a more ionized plasma. Unlike the last two stars, \stardis\ and \textsc{korg} show powerful agreement across every window in the spectrum. H$\alpha$ in particular is almost indistinguishable between the two codes, though \stardis\ does predict very slightly wider Lorentzian wings once again which were not present in the solar comparison. However, in part because the C$_2$ feature is absent in the \ion{Mg}{1} window due to the high temperature and low metallicity of the star, the two codes produce strikingly similar spectra in general. The one emergent discrepancy appears in the rightmost feature of the \ion{Ca}{2} triplet. \stardis\ shows an additional blended absorption feature around 8548 \AA that is absent in \textrm{korg}. 
%Should write more about this but not sure about it yet.

% We see that \stardis\ often predicts similar Doppler cores to \textsc{korg}, but slightly broader Lorentzian wings which are mostly seen in the strongest lines in each spectrum. Furthermore, the majority of lines predicted by \stardis\ appear slightly stronger than those predicted by \textsc{korg}. Because the majority of the difference appears in the wings, similar to HD122563 and the discussion in Section \ref{ss:4500K}, this primarily suggests that the Lorentzian broadening constants are scaled differently due to different plasma or \ion{H}{1} number densities. Overall, the codes agree quite well with discrepancies mostly appearing in the exact shape of the Lorentzians, and largely agreeing to better than 2\%.

\subsection{Continuum Comparison} \label{ss:continuum}

\begin{figure*}[t] 
    \centering
    \includegraphics[width=.45\textwidth]{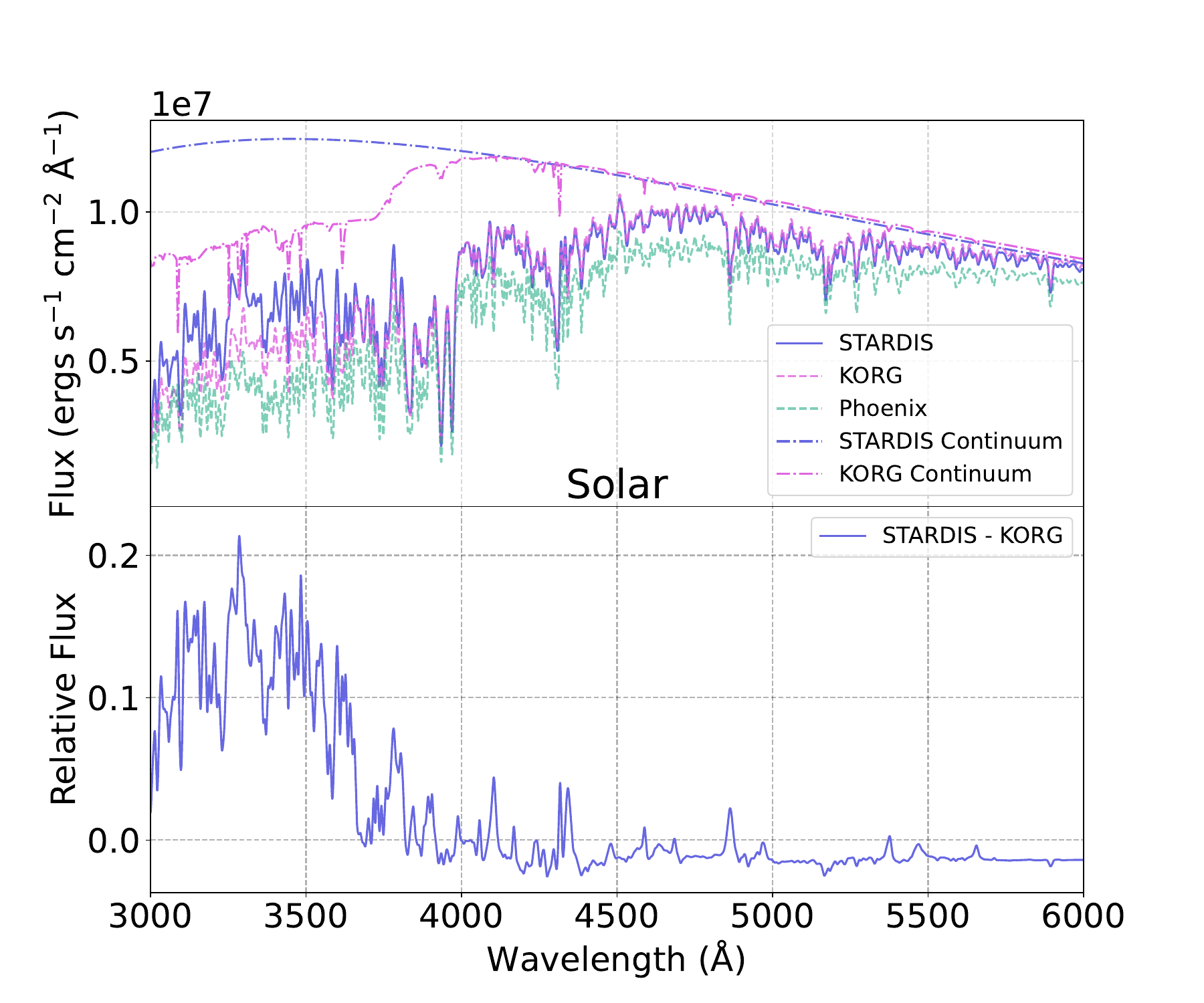}
    \includegraphics[width=.45\textwidth]{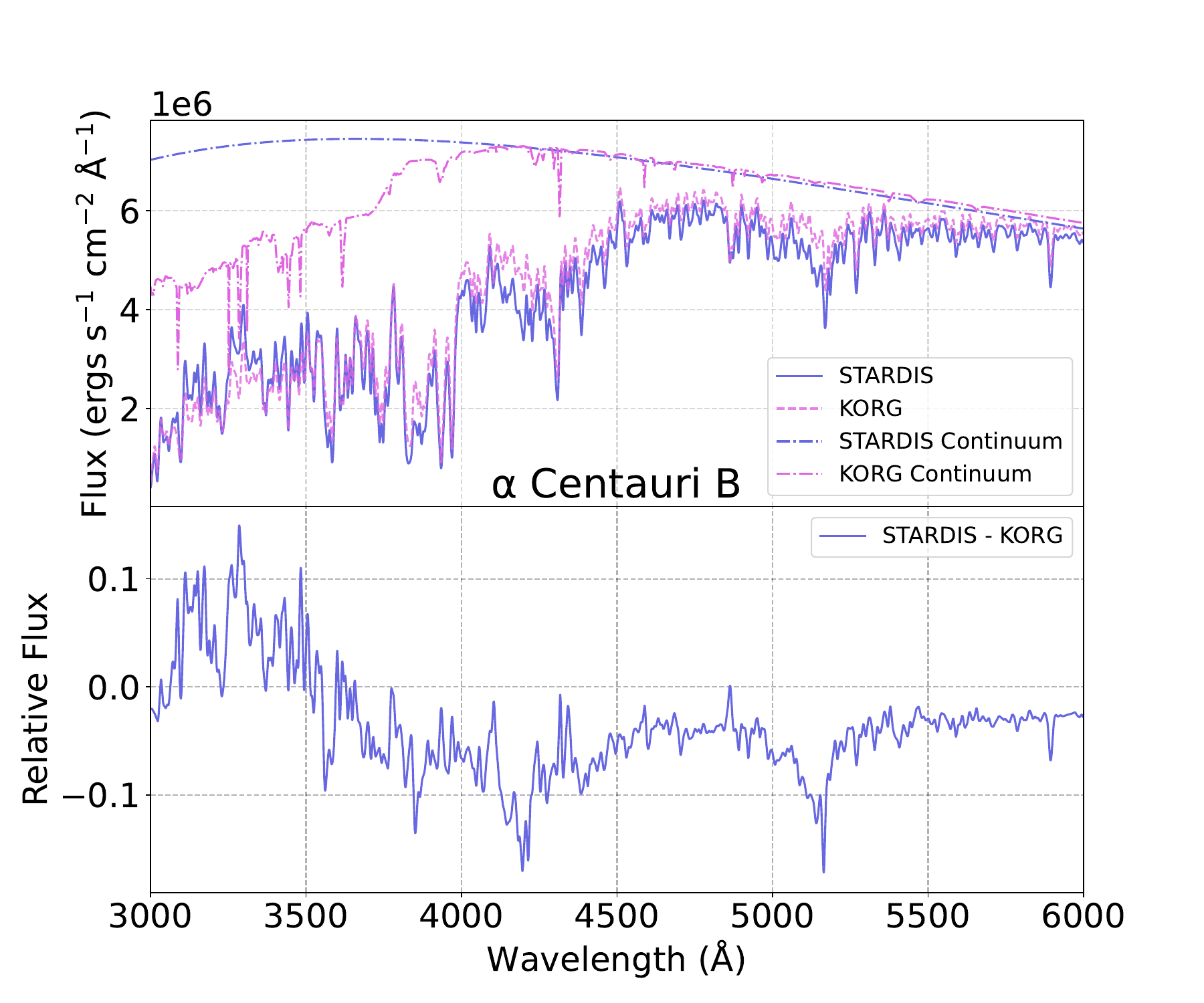}
    \includegraphics[width=.45\textwidth]{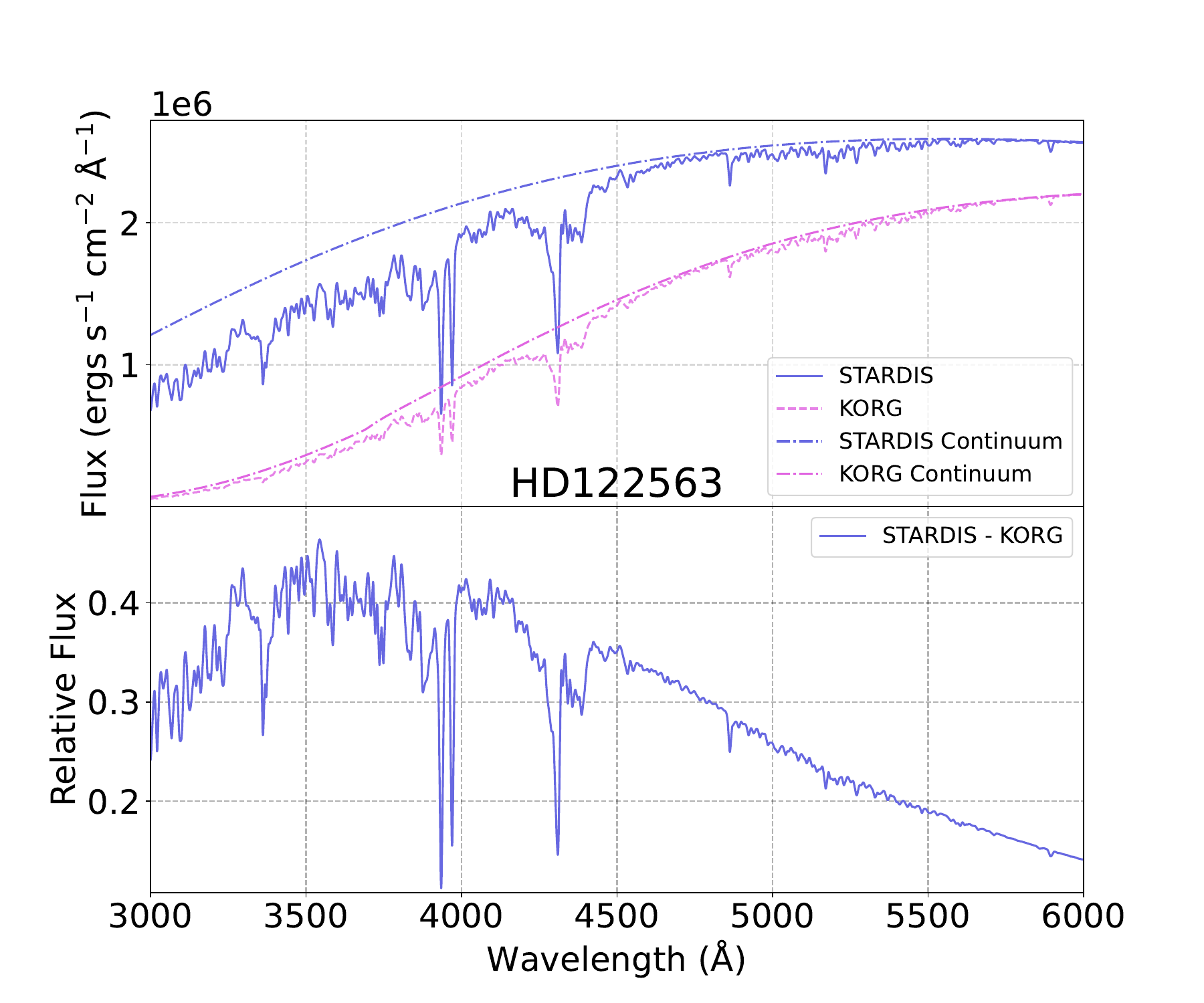}
    \includegraphics[width=.45\textwidth]{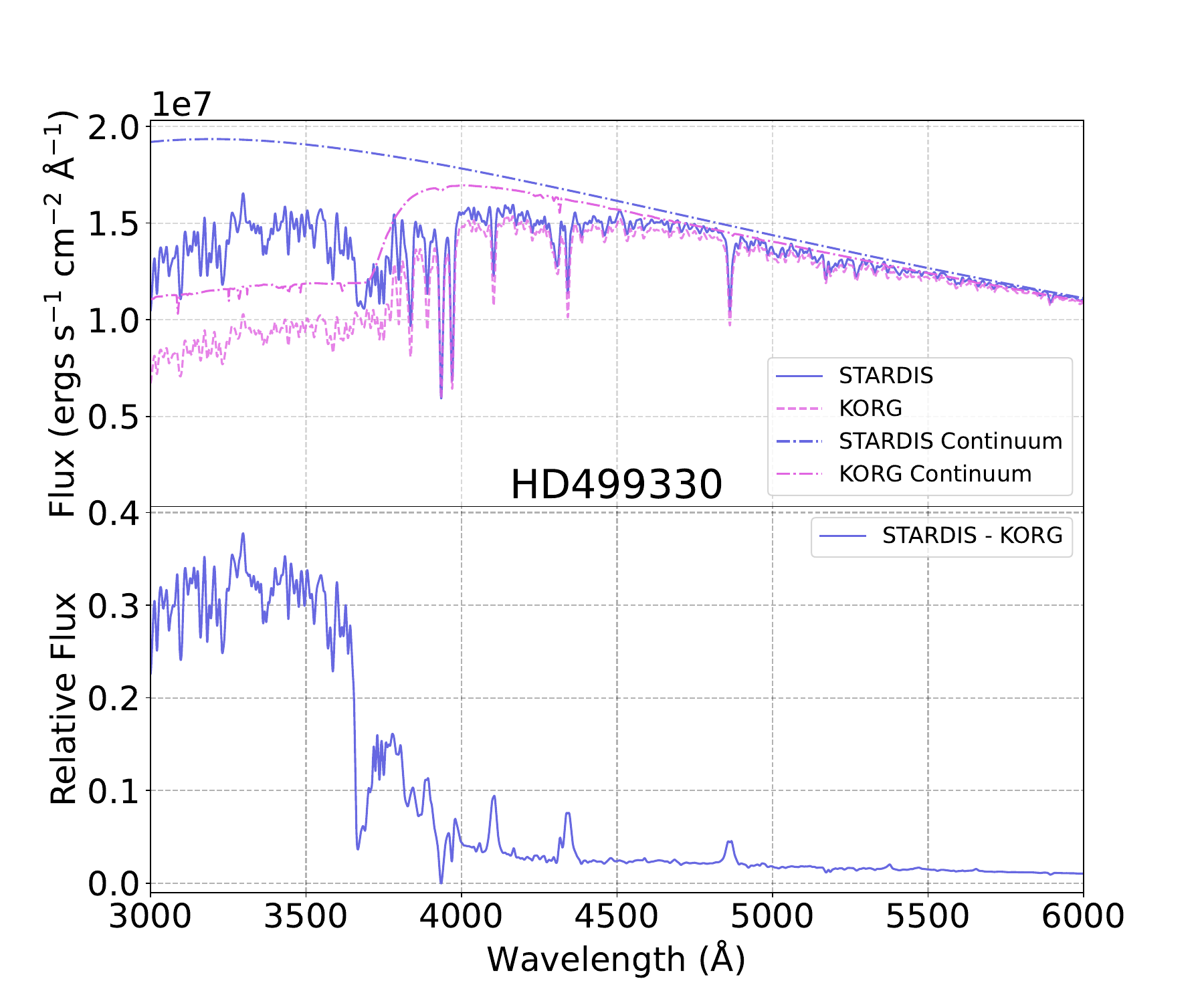}    
    \caption{A direct comparison of the Ultraviolet continuum differences between the four stars in Figures~\ref{fig:sol_comp}--\ref{fig:6250_comp}. The \stardis\ continuum starts to disagree heavily with \textsc{korg} bluer than 4000~\AA\ due to a current lack of high energy continuum opacity sources. }
    \label{fig:cntm_comp}
\end{figure*}

Finally, we show a comparison of absolute synthesized spectra in Figure~\ref{fig:cntm_comp}, focusing on green and near UV wavelengths and additionally showing the uncalibrated continuum of \stardis\ and \textsc{korg}. The figure highlights many of the trends we saw in our previous comparisons, where the Sun shows strong agreement between codes, and all stars agree far more in the redder than bluer wavelengths, primarily due to the absence of certain sources of continuum opacity that become significant bluer than 4000~\AA. HD122563, the metal poor red giant, shows the largest discrepancy. This effect could be the result of missing continuum opacities becoming more dominant in the extended atmosphere, but may also point to differences in solving the system in spherical geometry, which is unique to this star in the set. However, the lines themselves agree strongly, and continuum normalization is common practice in measuring line strengths and inferring the concentration of chemicals. 

\subsection{Closing Comparison Discussion} \label{ss:comparison_closing}

In redder wavelengths beyond 6000~\AA, the codes generally agree better than 3\%, with exceptions of strong molecular lines and small differences in the exact shape of line profiles, where a broader line can have significant disagreement in the wings of the line, just outside of the Gaussian core and at the bottom of the line. However, we note that the exact description of individual line profiles remains an open question in the community. 

There are two noteworthy differences in the comparison residual plots. First, individual spectral pixels (here sampled every 0.1~\AA) dominate the residual plots when somewhat different broadening profiles are predicted by the two codes. These differences vanish with broader wavelength bins or any sort of convolution, like those intrinsically part of any spectral observation and demonstrated in Figure~\ref{fig:solar_obs}. Second, \stardis\ and \textsc{korg} can disagree in the exact shape of the Lorentzian wings of lines that can be diagnosed in strong line features. 
 % \stardis\ provides the ability to calculate line profiles both with empirically calibrated line broadening parameters, and analytic prescriptions. 

% Disagreement between all existing stellar codes is substantial in the UV \citep[see e.g. ][]{wheeler_korg_2023}. 

% While these differences suggest the need for further investigation, we note that these differences are not completely unexpected. As the Sun is the most well understood star, it serves as a natural target calibrator for the development of any spectral synthesis code. 

%Caveats
% During this stage, the user can also choose to arbitrarily alter the chemical composition of the atmosphere. This is especially important to allow for parameter exploration and spectral fitting.
% % To allow for a tailored analysis, the chemical profile of the atmosphere, whether created originally from a source like MARCS or independently by the user, can be altered.
% However, it should be noted that the atmospheric structure can be affected by the chemical composition of that atmosphere through differences in the coupling of the radiation to the plasma (e.g., changes in radiation pressure on the plasma). Small departures from the chemical composition of the input atmosphere will not have significant effects on the physical atmosphere being modeled, but it is worth noting that \stardis\ will not verify the stability of the atmosphere requested. That is, \stardis\ will solve the radiative transfer equation for the atmosphere, whether or not the atmosphere being supplied is physically consistent or satisfies hydrostatic equilibrium. 

\section{Conclusions and Future Work} \label {sec:stardis_conclusions}

We present \stardis, a new 1D stellar spectral synthesis code. This code is written in Python, the language of choice of the astrophysics community. \stardis\ performs three major pieces of computation after ingesting necessary data from outside sources in regular operation. First, it solves the state of the plasma, detailed in Section~\ref{ss:stellar_plasma}. This mostly entails solving for the number densities of each relevant opacity source. With those number densities and the relevant plasma properties for each opacity source, the code then calculates the opacity of the plasma at each spectral pixel requested, as detailed in Section~\ref{ss:opacities}. Finally it solves the radiative transfer equation with a ray tracing prescription through the stellar atmosphere to obtain the emergent flux from the surface of the star, detailed in Section~\ref{ss:stellar_radiation_field}.

A comparison between \textsc{korg} and \textsc{phoenix} shows strong agreement, especially redder of the near-ultraviolet spectrum of the Sun. Between \textsc{korg} and \stardis\, normalized spectra agree on the level of 3\% or better, save for two exceptions. First, individual lines that are not yet incorporated in \stardis\ stand out as large residuals in the comparison as one would expect. Second, broadening differences in lines lead to individual spectral pixels that have large disagreement, usually resembling W shapes in the residuals. These disagreements, however, are generally on smaller scales than observationally distinguishable, and verifying which code, if either, is more physically accurate is beyond the scope of this work. 

\stardis\ is a part of the expanding and actively developed \textsc{tardis} codebase to encourage long-term sustainability and minimize the reproduction of shared processes (i.e., physics implementations and numerical solvers written share a common API). \stardis\ is entirely open-source and can be found at \href{https://github.com/tardis-sn/stardis}{https://github.com/tardis-sn/stardis}. Documentation, including installation and usage tutorials, can be found at \href{https://tardis-sn.github.io/stardis/}{https://tardis-sn.github.io/stardis/}.

% \stardis\ is built with modularity in mind, to be able to accurately model a wide variety of stellar systems that require different physics and approximations systems as needed. Being linked to the \textsc{carsus} atomic ingestion package also means that any new updates to atomic data can be immediately incorporated into simulations. 

% It is currently the only stellar spectral synthesis code known to the authors capable of directly ingesting MESA atmospheres. We advise caution when using \stardis\ to model MESA atmospheres as we have yet to comprehensively validate MESA models against known outputs, however we believe that the tool should enable scientists interested in unique systems to make new insights that were not yet possible. Furthermore, models can be tweaked at the Python object level to give the user as much flexibility as possible. This idea extends to the atomic data as well, being linked to the \textsc{carsus} atomic data ingestion package within the\textsc{tardis} codebase to allow for as flexible atomic data inputs as possible. 

% Currently \stardis\ has only been strongly validated on solar analogs. However, preliminary results are encouraging, and \stardis\ is currently able to produce spectra of sufficient accuracy to constrain stellar astrophysical parameters and chemical abundances. We soon intend to validate \stardis\ on a more comprehensive range of stellar types.

\stardis\ has several expansions and new physics planned or already in development. Detailed ionization-state and excited-level plasma solvers that will be used across \textsc{tardis} and \stardis\ are in development, and will allow for flexible investigation of departures from LTE and the effects that they have on stellar spectra and resulting composition measurements. These calculations will be particularly important in extending \stardis\ to the modeling of hot, massive stars, but could also allow for investigation of the solar corona.
The implementation of metal continuum opacities will allow for accurate spectral modeling in the UV. Finally, continued optimizations will further improve the speed of the \stardis\ code. We discuss some details of both performance and code implementation choices in Appendix~\ref{sec:benchmarks}. Our preliminary investigation suggests that the most computationally expensive parts of the code are good candidates for GPU implementations which are well supported in Python and will provide a massive reduction to code runtimes in large simulations.

% In the long term, we intend \stardis\ to be approachable without compromising on efficiency or precision. 
% In addition to the planned and in progress extensions to the physical and computational aspects of the code, we are currently producing comprehensive tutorials to showcase the various physics and computation schemes implemented in \stardis. 
We intend this code to provide a suitable tool to study the wide variety of stellar systems of interest across astrophysics, and that the approachability of the code will allow for more efficient and focused stellar research. 

\section{Acknowledgments}
The authors would like to thank Sean Couch, Michael Zingale, and Anirban Dutta for the scientific input, helpful discussion, and assistance in code development and funding acquisition.
The development of \stardis\ received support from GitHub and the Google Summer of Code initiative.
J.V.S is supported by NSF OAC-2311323 and DOE No. DE-SC0017955.
A.G.F. is supported by NSF OAC-2311323.
T.M.D.P. is supported by the Research Council of Norway through its Centers of Excellence scheme, project number 262622.
J.L. is supported by NSF-2206523 and DOE No. DE-SC0017955. C.J.F. is supported by the U.S. Department of Energy through the Los Alamos National Laboratory, which is operated by Triad National Security, LLC, for the National Nuclear Security Administration (Contract No. 89233218CNA000001).

\software{Astropy\footnote{\url{https://www.astropy.org/}} \citep{astropy_collaboration_astropy_2013, astropy_collaboration_astropy_2018, astropy_collaboration_astropy_2022}, 
pandas\footnote{\url{https://pandas.pydata.org/}} \citep{pandas_paper,pandas_software},
Scipy\footnote{\url{https://scipy.org/}} \citep{virtanen_scipy_2020},
NumPy\footnote{\url{https://numpy.org/}}\citep{harris_array_2020},
\textsc{tardis}\footnote{\url{https://github.com/tardis-sn/tardis}} (\citealt{kerzendorf_spectral_2014}; version: \citealt{kerzendorf_tardis-sntardis_2024})}



\bibliography{references}{}
\bibliographystyle{aasjournal}

\section{Contributor Roles}

\begin{itemize}
\item{Conceptualization: Joshua V. Shields, Wolfgang Kerzendorf}
\item{Data curation: Joshua V. Shields}
\item{Formal Analysis: Joshua V. Shields}
\item{Funding acquisition: Wolfgang Kerzendorf}
\item{Investigation: Joshua V. Shields}
\item{Methodology: Joshua V. Shields}
\item{Project administration: Wolfgang Kerzendorf}
\item{Software: Joshua V. Shields, Wolfgang Kerzendorf, Isaac Smith, Ryan Groneck, Andrew Fullard, Jaladh Signhal, Christian Vogl}
\item{Supervision: Wolfgang Kerzendorf, Joshua V. Shields} 
\item{Validation: Joshua V. Shields}
\item{Visualization: Joshua V. Shields}
\item{Writing - original draft: Joshua V. Shields}
\item{Writing - review \& editing: Joshua V. Shields, Wolfgang Kerzendorf, Andrew G. Fullard, Tiago M. D. Pereira, Christian Vogl, Jing Lu}

\end{itemize}

\appendix
\section{Benchmarks} \label{sec:benchmarks}

%Move this

\begin{table*}[ht]
    \centering
    \begin{tabular}{c|c|c|c}
           Simulation Width & \stardis, Serial (s) & \stardis\ Parallel, 8 threads (s) & \textsc{korg} (s) \\
          \hline
          50 \AA\  & 2.31 $\pm$ 0.006 & 1.6 $\pm$ 0.005 & 0.65 \\
          1000 \AA\  & 14.4 $\pm$ 0.021 & 6.98 $\pm$ 0.018 & 1.90 \\

    \end{tabular}
    \caption{Compared to \textsc{korg}, the only other recently developed stellar spectral synthesis code, \stardis\ is roughly two to ten times slower in most cases. However, we note that all times listed here are post-compile times. \stardis\ has a flat compile time of roughly 5 seconds, while \textsc{korg}'s compile time is close to 20 seconds in our testing. This makes \stardis\ significantly faster to compute a first spectrum for many cases, but slower in most other cases.}
    \label{tab:benchmarks}
\end{table*}

We show a comparison of code run-times in Table~\ref{tab:benchmarks} for two sets of realistic simulations using the same simulation inputs (i.e., the same line lists and spectral grids applied to a solar-type model). All testing was done on an AMD EPYC 7552 48-Core Processor on Ubuntu 20.04.6 LTS. \stardis\ is marginally but noticeably slower because a significant portion of the code is written in pure Python, while \textsc{korg} is written in Julia. However, \stardis\ achieves speeds comparable to  \textsc{korg} in large part because the computationally expensive parts of the code are compiled just-in-time with \textsc{numba} \citep{lam_numba_2015}. \textsc{numba} achieves speeds on par with C \citep{nikolaos_ziogas_productivity_2021} and is designed to be portable across different architectures, providing both parallelization and GPU support. \stardis\ includes parallelization for the most computationally expensive parts of the code in this way. Finally, we note that both codes need to compile for first execution, but \textsc{korg}'s compile time is about four times as long ($\sim$ 5 seconds vs. $\sim$ 20 seconds) which makes \stardis\ significantly faster to first spectrum.

\end{document}